\documentclass[prb,twocolumn,nofootinbib,longbibliography,superscriptaddress]{revtex4-2}
\usepackage[utf8]{inputenc}
\usepackage{graphicx}
\usepackage{lipsum}
\usepackage{multirow}
\usepackage[normalem]{ulem}
\usepackage{braket}
\usepackage{textgreek}
\usepackage{epsfig}
\usepackage{floatrow}
\usepackage{amssymb,amsmath,amsfonts,hyperref}
\usepackage{wasysym}
\usepackage{latexsym}
\usepackage{xcolor}
\definecolor{je}{rgb}{0.858, 0.188, 0.478}
\definecolor{so}{rgb}{0.19,0.32,1.52}
\usepackage{verbatim}
\usepackage[caption=false]{subfig}
\usepackage{braket}
\usepackage{lipsum}
\usepackage{cleveref}
\newcommand{\la}{\langle}
\newcommand{\ra}{\rangle}

\newcommand{\cd}{\partial}

\newcommand{\e}{{\mathrm{eff}}}

\newcommand{\Gc}{\mathcal{G}}
\newcommand{\Sc}{\mathcal{S}}
\newcommand{\Hc}{\mathcal{H}}
\newcommand{\Pcs}{\mathcal{P}_{\mathrm sh}}

\newcommand{\oo}[1]{\overline{\overline{#1}}}
\newcommand{\ovl}[1]{\overline{#1}}

\newcommand{\AB}{{\texorpdfstring{\text{AB}}{ab}}}
\newcommand{\ABB}{{\texorpdfstring{\text{AB$^*$}}{abb}}}

\begin{document}

\title{Discrete scale invariant fixed point 
 in a  quasiperiodic classical dimer model}
\author{Sounak Biswas}
\affiliation{Institut für Theoretische Physik und Astrophysik, Universiät Würzburg, 97074 Würzburg, Germany}
\affiliation{Rudolf Peierls  Centre  for  Theoretical  Physics, Parks Road, Oxford  OX1  3PU,  United  Kingdom}
\author{S. A. Parameswaran}
\affiliation{Rudolf Peierls  Centre  for  Theoretical  Physics, Parks Road, Oxford  OX1  3PU,  United  Kingdom}

\begin{abstract}
  We study close-packed dimers on the quasiperiodic Ammann-Beenker (AB) graph, that was recently shown to have the unusual feature that  hard-core dimer constraints are \textit{exactly} reproduced at successive discrete length scales. This observation led to a conjecture  that it would be possible to construct an exact real-space decimation scheme where each iteration preserves both the quasiperiodic tiling structure and the constraint. 
  Here, we  confirm this conjecture by explicitly constructing  the corresponding  renormalization group transformation and show, using large-scale Monte Carlo simulations, that the dimer distributions flow to a fixed point with  non-zero dimer potentials. We use the fixed-point Hamiltonian to demonstrate the existence of slowly decaying dimer correlations. We thus identify a remarkable example of a classical statistical mechanical model whose  properties are controlled by the fixed point of an exact  renormalization group procedure exhibiting discrete scale invariance but lacking translational and continuous rotational symmetries.
\end{abstract}
\maketitle
\section{Introduction}
\begin{figure}[t]
    \includegraphics[width=0.8\columnwidth]{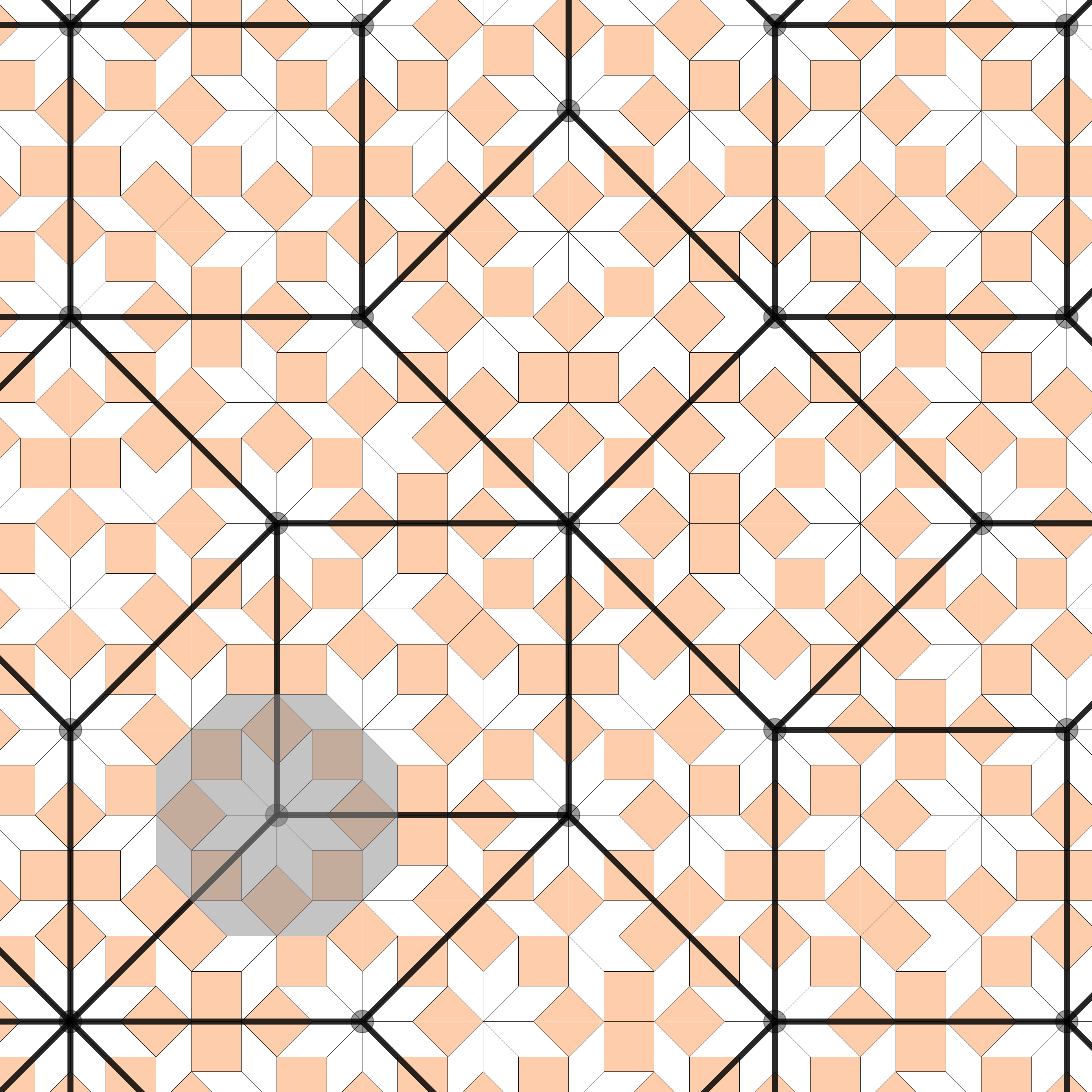}
    \caption{A finite patch of the quasiperiodic Ammann-Beenker tiling (or graph). The tiling is built out of two kinds of plaquettes: a square and a rhomb. The plaquettes have the same edge length; the rhomb has internal angles  $3\pi/4$ and $\pi/4$. The tiling has no symmetries of translation. The grey shaded region shows the local empire of an 8-vertex: the set of tiles which always simply connected to an 8-vertex in the tiling}
    \label{fig:abpatch}
\end{figure}

Scale invariance is a striking  phenomenon that appears in various guises across a  range of problems in physics and beyond~\cite{Cardy_book,Barenblatt_book}.  Perhaps its most  familiar setting is in the context of critical phenomena, where it is one of the hallmarks of continuous phase transitions, both classical and quantum~\cite{SKMa,Goldenfeld,Sachdev}. In the standard lore, critical fluctuations are correlated on all length scales, making their long-wavelength asymptotic properties --- that average over many correlated degrees of freedom ---  insensitive to microscopic details. This leads to striking universality across superficially very different problems.  Similar arguments also apply to stable critical phases that lack long-range order, as in the two-dimensional XY model. This intuitive  link between scale invariance and universality can be made more quantitative  via the renormalization group (RG)~\cite{WilsonKogut,Goldenfeld,Cardy_book}. The RG describes a class of iterative procedures for  increasing the length scale over which a system is studied, while discarding all but the essential data about smaller scales.  Fixed points of such rescaling procedures that describe the universal behaviour of critical points or stable critical phases usually exhibit continuous scale invariance, signalled by the divergence of the correlation length, i.e. the distance beyond which correlations become statistically independent.

An implicit assumption in nearly all RG approaches and in the preceding discussion is that as a lattice model is coarse-grained, its microscopic details cease to matter and therefore the emergent scale invariance is equivalent to that of a continuum field theory. Indeed, often one simply asserts that a continuum description of the fixed point exists, and then justifies the neglect of lattice-scale effects \textit{a posteriori} by showing that, in RG parlance, they are irrelevant perturbations, i.e. shrink under successive iterations of the RG. Long-wavelength properties controlled by such a continuum fixed point exhibit \textit{continuous} rotational and translational symmetry, and are invariant under \textit{arbitrary} rescalings of the  coordinates. Often these combine to generate invariance under a larger group of conformal transformations, with particular significance in two dimensions~\cite{BPZ,Ginsparg,Cardy_LesHouches}.

In this work we identify a rare exception to these general rules: namely, a statistical mechanical problem that admits an exact RG fixed point that lacks translational symmetry, and is only invariant under discrete point-group rotations. Most strikingly, however, the fixed point only has \textit{discrete} scale invariance (DSI), under rescaling distances by an even power $\delta^{2n}$ of the silver mean $\delta=1+\sqrt{2}$. This is in striking contrast to the behaviour of RG fixed points of continuum field theories.

The new fixed point combines two ingredients: quasiperiodicity and constraints. We study a model of close-packed dimers~\cite{Kasteleyn61,Kenyon_review} on the quasiperiodic Ammann-Beenker (AB) tiling~\cite{GrunbaumShephard,BaakeGrimm}, (Fig.~\ref{fig:abpatch}) with the usual constraint that exactly one dimer touches each site\footnote{On finite geometries the constraint may be modified to allow a single unmatched site, as we clarify below.}, defining a so-called {\it maximum} dimer covering. (In the graph theory literature, dimer coverings are often termed [vertex] matchings~\cite{LovaszPlummer} since each dimer uniquely matches a pair of vertices; we will use these terms interchangeably below.) Constrained models resist a naive application of the RG, since hard constraints usually soften on coarse-graining. Hence such problems are usually tackled by reformulating them in terms of new variables that automatically satisfy the constraint. For example,  two-dimensional dimer models on periodic lattices can be mapped to ``height models'' that can then be studied by taking a continuum limit~\cite{Blote_Hilborst,Nienhuis_Hilborst_Blote,Henley97,Moessner_Raman}. Similarly,  quasiperiodicity of the lattice is often an irrelevant perturbation to a continuum fixed point~\cite{Luck}: intuitively, the distinction between periodic and quasiperiodic structures is averaged out at the longest length scales. (This can be formalized by considering the    `wandering exponent' of averages over quasiperiodic regions.)  Surprisingly, however, the \textit{combination} of  quasiperiodicity and constraints resists such simplifications. The quasiperiodic structure forces the height description to have a complicated nonlocal free energy. As we show below, the constraint in turn forces the RG to retain information about the inflation properties of the underlying quasiperiodic tiling, with DSI as a natural consequence. 

The fixed point we identify in this paper cannot be readily accessed from a continuum limit, since  such theories  usually have continuous scale and translational symmetries. Hence the RG scheme must work directly with lattice variables. Consequently, our approach is rooted in an older yet arguably more physically transparent RG procedure: the ``block spin'' decimation scheme originally proposed by Kadanoff and Migdal in the analysis of local spin models~\cite{Kadanoff_RG,Migdal_Rg,KadanoffRMP}. Similarly to the block-spin scheme, we work directly in real space, and in each iteration of the RG identify subsets of degrees of freedom (dimers) that define new degrees of freedom at the next scale. The choice of scale transformation $x\to \delta^2 x$ in each step is dictated by the discrete scale symmetry encoded in the ``inflation rule'' that generates the AB tiling\footnote{The tiling is self-similar under a pair of inflations.}. Remarkably, our RG transformation preserves the single-dimer-per site constraint  \textit{exactly} at each successive scale, leading to a direct RG map between dimer models at successive length scales. The well-known failure of the block-spin approach to eliminate lattice-scale information now emerges as a feature, since this enables the retention of quasiperiodic structure  inherent to the fixed point.  We circumvent the technical challenges usually encountered in implementing real-space blocking transformations by using large-scale efficient Monte Carlo sampling of dimer correlations to implement the RG transformation explicitly. This strategy leads to a huge variety of algorithms varying in the specifics of the choice and implementation of the RG transformation,  and are often collectively referred to as the Monte Carlo Renormalisation Group (MCRG) methods~\cite{Swendsen_PRL,VarMCRG,NeuralMCRG}.  

The self-similar fractal properties of quasiperiodic systems are well-known: for instance, the inflation procedure at the root of our RG decimation is one example of such self-similarity. However, the self-similarity of the  dimer model defined on the quasiperiodic AB tiling lies beyond this: requiring that the dimer partition function reproduce itself, including constraints, under discrete scale transformations is highly nontrivial.  DSI at the level of the partition function does not emerge in other problems: for example, implementing a real-space RG transformation on  Ising and Potts ferromagnets on quasiperiodic tilings do {\it not} lead to a DSI fixed point~\cite{penrose_rsrg,ab_rsrg}. Lattice models where  exact DSI emerges are typically defined on cousins of the ``Bethe lattice''~\cite{Kaufman_Griffiths1}, with  tree-like hierarchical structures that cannot be embedded in any finite dimension~\cite{MeuriceEA, Derrida1983,Derrida1984}. This is clearly in contrast to the quasiperiodic AB graph, which is constructed from a tiling of the plane. Classical lattice statistical mechanics problems with DSI fixed points are rather unusual; the only other example we are aware of is from a recent work on DSI percolation transition~\cite{SommersEA}.

Although one aspect of our RG transformation --- the exact reproduction of the dimer constraint at successive RG scales --- was motivated by graph theoretic considerations in a recent paper by us and others~\cite{LloydEA}, our construction of an explicit RG transformation and the determination of the fixed-point Hamiltonian and its correlation structure represent a substantial advance, confirming and substantially expanding on the ideas proposed in that previous work.    
The fixed-point structure we identify and explore  represents a striking departure from the known universality classes studied in classical statistical mechanics. Parallel and complementary work using machine-learning methods to identify the effective degrees of freedom discover them to be hardcore dimers without using any prior information about the model~\cite{GoekmenEA2}. While this feature is built into our RG transformation by hand,  this enables us to explicitly calculate effective Hamiltonians, characterise fixed points, and study critical phenomena associated with our model.

The rest of this paper is organized as follows. In section ~\ref{sec:review}, we
briefly introduce the dimer model on the Ammann-Beenker tiling and review how
the effective dimer constraint emerges at all scales. In Section
~\ref{sec:rgtrans}, we explicitly construct a Renormalisation Group
transformation to coarse-grain the dimer problem on the AB graph. We implement
this RG transformation with the aid of extensive Monte Carlo simulations,  and
show that the distribution of effective dimers thus obtained reaches a fixed point under the RG transformation.  In Sec.~\ref{sec:effham}, we calculate the
effective Hamiltonian describing the coarse-grained dimers. We show that the
couplings  flow to a fixed point under our RG transformation, establishing a
fixed point with DSI. This enables us to explicitly write
down the fixed point Hamiltonian controlling the dimer model on the AB tiling.
In Sec.~\ref{sec:results}, we investigate the properties of the fixed point Hamiltonian.  We first explore the consequences of the fixed point theory on long-range connected correlations of dimers. We show that 
dimer correlation functions are consistent with power-laws with log-periodic modulations, which signals a critical 
point with DSI. Finally we investigate the loop ensemble defined by the overlap graphs of two decoupled  dimer models; 
we show conclusive numerical evidence of the these loops being critical, albeit with discrete, instead of continuous, scale-invariance.
We calculate the critical exponents associated with this loop-ensemble. In Sec.~\ref{sec:discussion}, we summarise our progress and outline some 
future directions directly motivated by our results.

\section{Dimers on the AB tiling}
\label{sec:review}
We begin with a brief introduction to the properties of AB tilings. We also review the results of  Ref.~\cite{LloydEA} on the dimer problem on the AB tiling, to motivate the RG perspective adopted in the balance of the paper.

\subsection{AB graphs and discrete scale symmetry}

\begin{figure}
    \includegraphics[width=0.7\columnwidth]{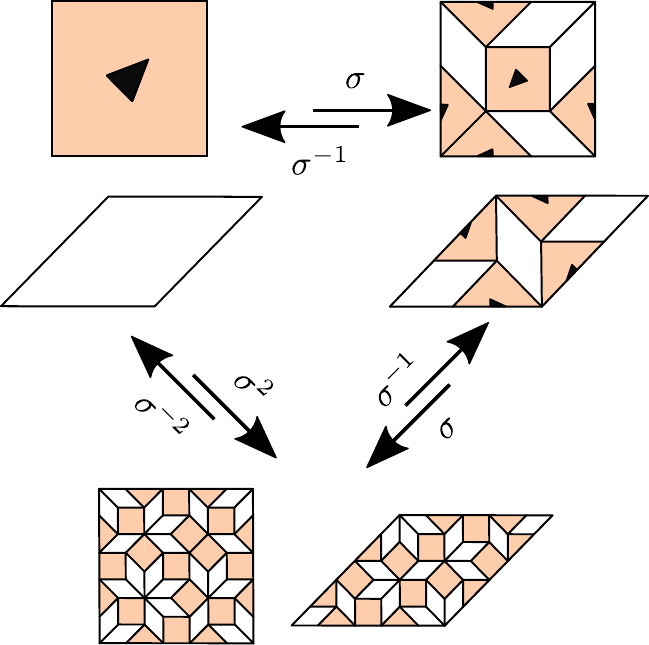}
    \caption{Inflation rules $\sigma$ which generate the AB tilings. Given each square or rhomb tile, $\sigma$ first decomposes each tile into a smaller set of tiles as shown, followed by rescaling so that the new tiles are of the same size as before. One can also consider the inverse process $\sigma^{-1}$, called deflation which involves composing the smaller tiles back to form the larger tiles, followed by similar rescaling. }
    \label{fig:inflation}
\end{figure}
\begin{figure*}
  \includegraphics[width=7cm]{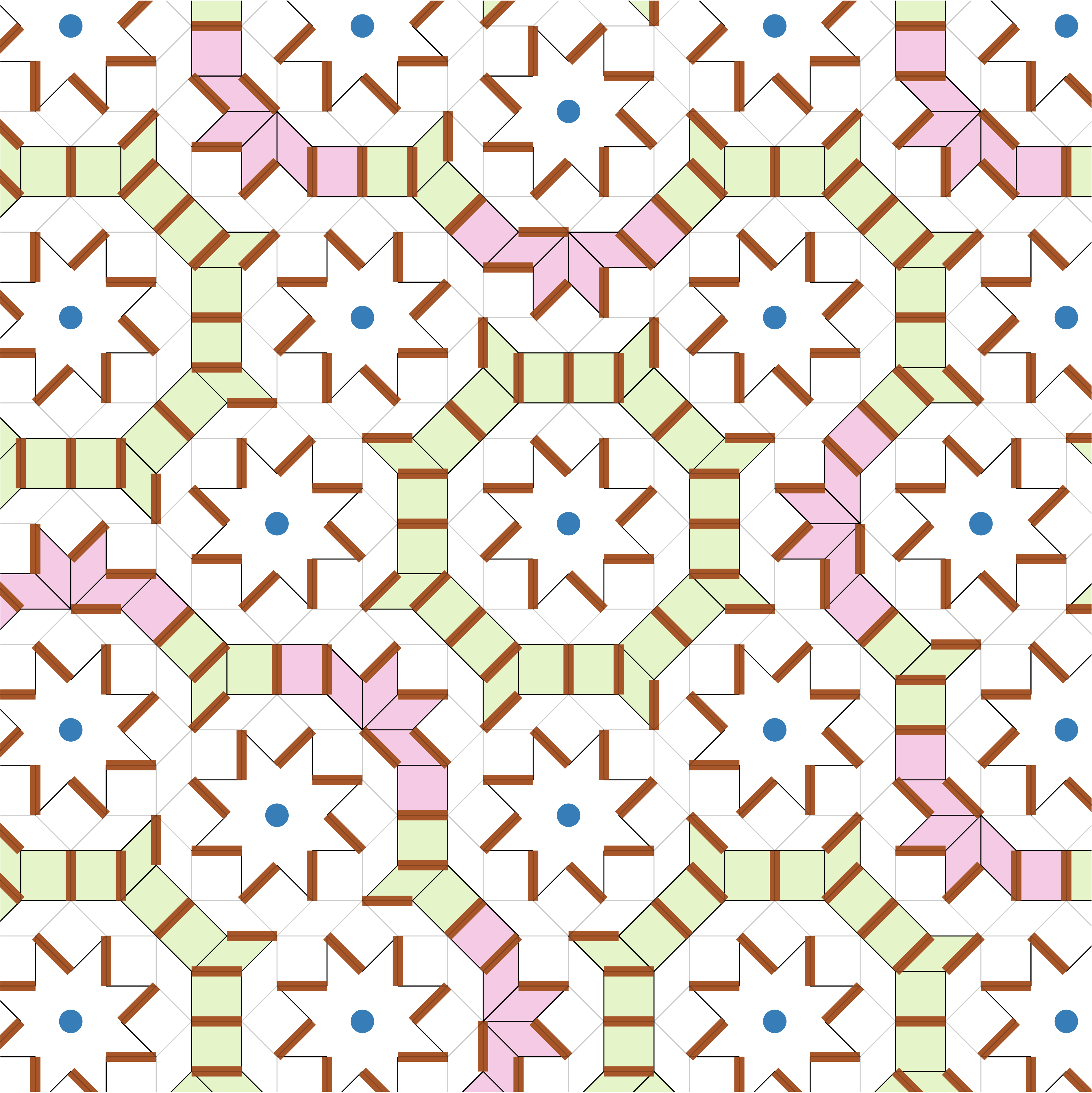} \hspace{1cm}
    \includegraphics[width=7cm]{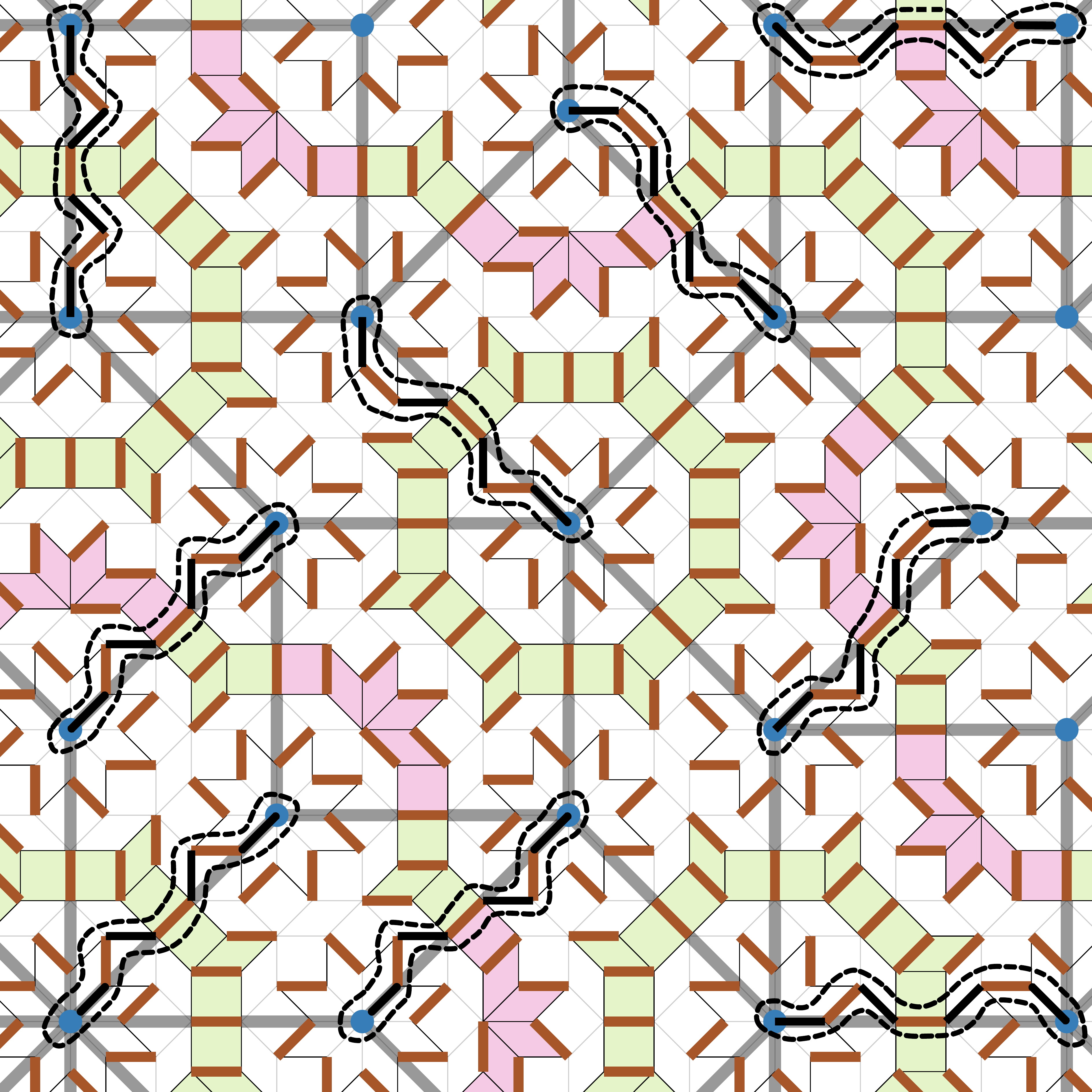}
    \caption{\emph{Left}: The \ABB~graph is obtained from the \AB\ graph by removing all the 8-vertices (marked by solid blue circles). The vertices in the \ABB\ graph can be partitioned into mutually exclusive sets: loops of 16 sites called the stars, and quasi-1D objects which we call ladders. The stars can be matched easily. The ladders can be further decomposed into two kinds of segments (they  are shaded with pink and green). By matching each of these segments independently, the whole \ABB~graph can be matched up. \emph{Right}: Once the \ABB\ graph is matched up, it remains to match up the 8-vertices by pairing them up with alternating paths. The 8-vertices lie on the positions of another, larger, \AB\ graph with edge lengths larger by a factor of $\delta_s^2$; the edges of this graph are shown by thick grey lines. It can be shown that pairs of monomers corresponding to edges of the larger \AB\ graph can always be matched up by augmenting alternating paths between them; thereby reducing the problem to that of matching up vertices of the larger \AB\ tiling. This can be iterated repeatedly to match up the whole \AB\ graph, hinting at possible effective matching problems of \AB\ graphs at successive lengthscales.  }
    \label{fig:starsandladders}
\end{figure*}

  The \AB~tiling is a quasiperiodic tesselation of the plane by square and rhombus plaquettes. The sides and corners of these plaquettes define the edges and vertices of a quasiperiodic graph, that will be the focus of our study. (Henceforth we use graph and tiling interchangeably; the meaning  will be clear from context.)
  Although there are many different prescriptions to generate quasiperiodic tilings, we focus on the approach where DSI is most clearly evident:  ``inflation'', a procedure that  generates a larger quasiperiodic tiling from a given ``seed'' tiling. Consider an initial such seed, denoted $\Gc^0$. The ``inflation map'' $\sigma$ provides a rule for decorating each tile of $\Gc^0$ with new edges and vertices (and deleting some edges), followed by  a spatial rescaling by a factor $\delta$, such that the area of each distinct tile type is preserved, generating a larger tiling $\Gc^1=\sigma(\Gc^0)$. Continuing this process iteratively generates an \AB~tiling of the infinite 2D plane as the number of iterations, $N\to \infty$. 
  
  In the $N\to \infty$ limit, a single inflation transformation $\sigma$ maps a tiling to a $\pi/2$ rotated version of itself scaled by a factor $\delta$. Applying the transformation twice, i.e $\sigma^2$ maps a tiling to an identically oriented one that is bigger by a factor $\delta^2$. Evidently, the existence of this inflation procedure demonstrates DSI of the quasiperiodic tiling.

Since the RG ethos is to eliminate degrees of freedom, it is more natural to consider the inverse ``deflation'' maps $\sigma^{-1}$ and $\sigma^{-2}$, which shrink the tiling.  While both $\sigma^{-1}$ and $\sigma^{-2}$  generate DSI, we will focus on the latter, as it leads to a more natural RG rule due to the behaviour of vertices under this transformation, which we now describe.

To understand the special properties of $\sigma^{-2}$, first observe that   ~\AB~tilings have 7 ``types" of vertices, each associated with a specific configuration of tiles 
  surrounding it: $3$, $4$, $5_A$, $5_B$, $6$, $7$ and $8$, the names reflecting the coordination numbers of the vertices. All vertices of a specific type develop a specific neighbourhood of tiles under inflation ($5_A$ and $5_B$ imply the development of different neighbourhoods of tiles). Crucially, the action of $\sigma$ changes the type of existing vertices while also adding new vertices and removing some edges. Under the action of $\sigma^2$, vertices of all coordination numbers on the original tiling $\Gc_0$ map to $8$-fold coordinate vertices (henceforth, $8$-vertices) of the double-inflated tiling $\sigma^{2}(\Gc_0)$. Inverting the map, an immediate corollary is that only $8$-vertices survive two deflations. In other words, given an \AB~tiling $\Gc$, the 8-vertices lie on the vertices of another \AB~tiling $\sigma^{-2}(\Gc)$,
whose  lengths are larger by a factor of $\delta_s^2$. We can therefore define a ``decimation'' transformation  where removing all vertices but the 8-vertices
gives us the same tiling rescaled by $\delta_s^2$. This is the decimation we 
will use to define our RG transformation in the next section to generate
effective Hamiltonians on a sequence of decimated graphs  with lengthscales given
by integer powers of $\delta_{s}^2$. This RG transformation is naturally defined with the double deflation $\sigma^{-2}$, rather than the single deflation $\sigma^{-1}$ for which no comparable simplification of the vertex mapping is possible.

Since  8-vertices will play an important role in our discussion, it is helpful to introduce some additional nomenclature relating to them. We mentioned that only $8$ vertices survive two deflations;  we can generalize this idea to define an order-$n$ 8-vertex, or an $8_n$ vertex,
  as an $8$ vertex which survives exactly $n+2$ deflations, remaining an $8$-vertex for $n$ of these deflations. $8_n$-vertices map to  $8_0$ vertices under $n$ deflations, while $8_0$-vertices map to  $8_n$ vertices under $n$ inflations. Like all vertices in a quasiperiodic tiling, 8-vertices are associated with a ``local empire", shown in Fig.~\ref{fig:abpatch}, comprising the set of
tiles which appear in the neighbourhood of \emph{each} 8-vertex, and are
simply connected to them.  We see that the local empire of the 8-vertex exhibits $D_8$-symmetry
associated with eightfold rotations. Since an $8_n$ vertex is obtained by
inflating an $8_0$-vertex $n$ times, and the inflations preserve the $D_8$
symmetry of the local empire, it is clear that an $8_n$ vertex has a $D_8$-symmetric local empire
whose radius scales as $\delta_s^n$.

\subsection{Structure of perfect dimer covers on AB graphs}

We now summarize some key properties of perfect dimer coverings or perfect matchings on AB graphs. Perfect matchings are configurations where every vertex participates in exactly one dimer. The existence of such covers on the \AB~tiling is nontrivial:  other quasiperiodic
graphs are known to host a finite density of monomers, i.e. vertices that participate in no dimers, even in  maximum dimer covers~\cite{Flicker_Simon_Parameswaran} (those with the largest possible number of dimers).  On periodic graphs that admit perfect dimer coverings, one can first construct one on a suitably chosen finite patch and then extend it to the whole graph by periodic repetition, but this construction is obstructed by quasiperiodicity. One must instead adopt different strategies; since the construction of perfect dimer coverings on \AB~tilings is central to the rest of out discussion, we now briefly review the arguments involved.

Let us begin with an infinite \AB~graph  $\Gc$ a for which a perfect matching is
to be constructed.  To proceed, we first consider an auxiliary graph, $\Gc^*$,
which is obtained from $\Gc$ by removing all the 8-vertices from it. (We will denote by \ABB~ graphs constructed by removing 8-vertices from
\AB~graphs.) A perfect dimer cover on $\Gc^{*}$ and other \ABB~graphs can be
constructed by first decomposing the vertices of the graph into non-overlapping
quasi-1D objects, termed stars and ladders, displayed in Fig.~\ref{fig:starsandladders}. The stars are
rings of 16 vertices which can be easily perfectly matched. The ladders are
aperiodically repeating sequences of two different kinds of segments (displayed
as plaquettes shaded with green and pink), and a perfect dimer cover on ladders
can be constructed by independently assigning perfect dimer covers to these
segments. Applying these  procedures on each star and ladder gives us a perfect matching on $\Gc^{*}$. It is possible to prove that in any perfect dimer cover on an \ABB~graph, all dimers lie entirely within
the ladders and the stars; there is never a dimer on the edges that link vertices
belonging to different stars or ladders.  This leads to a decoupling of the
dimer partition function on \ABB~ graphs into the partition functions on individual
ladders and stars, enabling transfer matrix computations of its statistical
mechanical properties~\cite{LloydEA}.   

\begin{figure}[t!]
  \includegraphics[width=8cm]{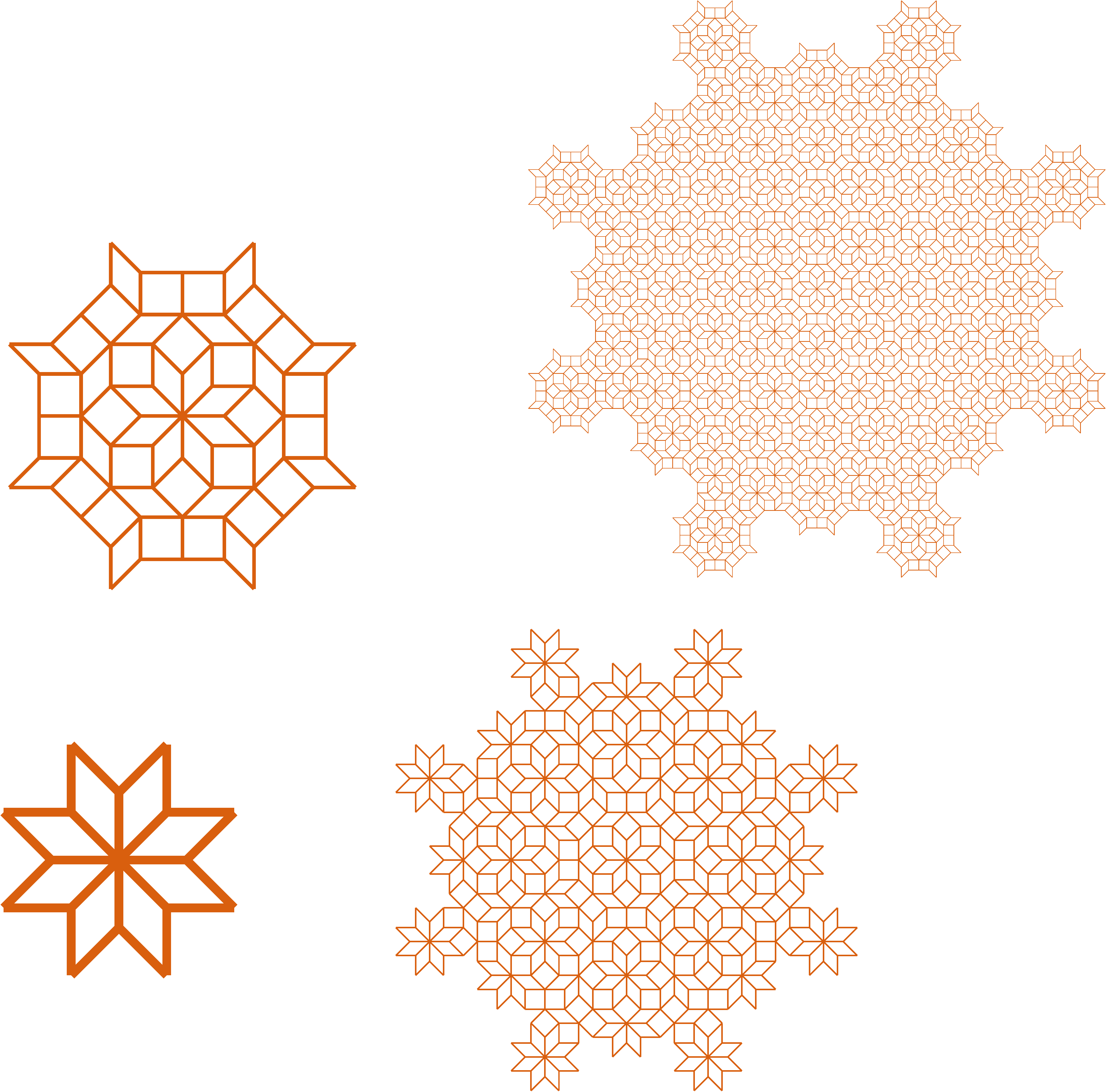}
  \caption{$8_n$-regions in \AB\ graph for $n=0,1,2,3$:  for each region, exactly one dimer connects vertices within the region to vertices outside. This imposes an effective dimer constraint on the region as a whole, making the $8_n$-regions effective vertices.}
  \label{fig:effective_vertices} 
  \end{figure}

Lifting a perfect matching from  the \ABB\ graph to the full \AB~graph results in an imperfect matching with a  monomer (i.e., a vertex with no dimers touching it) on each 8-vertex of \AB, since these were absent by construction on \ABB.
To construct a perfect matching on the \AB~ graph $\Gc$, we have to eliminate
these monomers. To do so, we must find ``alternating paths" between the two
monomers and ``augment" them. Here, an alternating path
between two vertices is a sequence of edges terminating at the two vertices such
that every alternate edge hosts a dimer; augmenting an alternating path between
monomers involves flipping the occupancy state of dimers on each edge on the
path, thereby annihilating the two monomers while increasing the number of
dimers by one. As explained above,
the 8-vertices of the \AB~graph $\Gc$ lie on the vertices of the twice deflated
graph $\sigma^{-2} (\Gc)$. It can be shown that starting from the perfect
matching in the auxiliary $\Gc^{*}$ graph, one can always annihilate monomers
at any pair of 8-vertices which corresponds to an edge of the twice-deflated
graph $\sigma^{-2}(\Gc)$(Fig.~\ref{fig:starsandladders}).  We are left with the task of annihilating these monomers pairwise, which is equivalent to the
problem of constructing a perfect matching of vertices (a dimer cover) on the graph $\sigma^{-2}(\Gc)$.  Thus, the
problem of constructing perfect matchings in $\Gc$ has been reduced to the
problem of constructing perfect matchings in $\sigma^{-2}(\Gc)$. This can be treated
exactly as $\Gc$ was treated, \textit{i.e.}, by first  matching the \ABB\ graph corresponding to $\sigma^{-2}(\Gc)$
leaving monomers on the $8$-vertices of $\sigma^{-2}(\Gc)$. Annihilating these monomers are now equivalent 
to the matching problem in $\sigma^{-4}(\Gc)$.
Iterating this
process gives us a perfect matching in the infinite tiling $\Gc$.
For finite patches of tiling, we can iterate this procedure until the graph
$\sigma^{-2n}(\Gc)$ has an $O(1)$ number of vertices, and then match them up
(possibly leaving behind a  $O(1)$ number of monomers depending on the
boundaries of the starting graph).

To understand this construction of perfect matchings, it is instructive to recall the notion of $8_n$ vertices, i.e. $8$-vertices which survive exactly $n+2$ deflations, as  defined in the preceding subsection. The iterative procedure outlined above matches $8$-vertices order by order:
the first iteration matches up everything except  8-vertices, the second iteration
matches all the remaining vertices except the $8$-vertices of $\sigma^{-2}(\Gc)$, \textit{i.e.}, the $8_0$- and $8_1$-vertices. The $n$-th iteration matches up $8_{n-2}$ and $8_{n-1}$-vertices with $8_k$-vertices for $k\geq n$ remaining unmatched, to be matched in subsequent iterations.

A consequence of this construction is that we can associate $D_8$-symmetric local empires of an $8_n$-vertex
 with ``effective vertices''. An effective vertex is a simply-connected region surrounding an $8$-vertex, such that there is exactly one dimer between vertices in the region and the rest of the graph. It is this one-dimer constraint on the region as a whole that leads us to consider it an effective vertex: if one coarse-grains this region to a single vertex, then the dimer constraint is reproduced exactly on the coarse-grained vertex.
The largest such region in the local empire of an $8_n$ vertex has radius $\sim \delta_s^{2n}$, and we refer to it as an  $8_n$-region. The structure of these $8_n$ regions are such that an $8_n$-region nests $8_k$ regions 
for all $k<n$. The effective vertices ($8_n$-regions) for the first few $n$ are displayed in Fig.~\ref{fig:effective_vertices}. 

For proofs of the above statements, which involve the use of ideas from bipartite matching theory, the reader is referred to Ref.~\onlinecite{LloydEA}. The matching theory approach  automatically singles out the stars and ladders of \ABB, and also allows for a clear proof of the effective vertex property based on the associated graph decompositions.

For our purposes, the central message to take from these considerations is that the one-dimer constraint on $8_n$ regions effectively reproduces the hard-core dimer
constraint at successive scales of coarse-graining. This remarkable feature strongly suggests that the dimer problem on an
\AB~tiling $\Gc$ can be recast as an effective dimer problems at each successive RG scales. Specifically, 
 the effective  dimer problem at  scale $\delta_s^{2n}$ concerns
$8_{2n}$  and $8_{2n-1}$ regions, which act as ``effective vertices" on
 the \AB~tiling $\sigma^{-2n}(\Gc)$. In the remainder of this paper, we make this intuition concrete by constructing a explicit RG
transformation and investigating the effective dimer models at different
scales.

\section{The RG transformation and the fixed point} 
\label{sec:rg}

We now present an explicit real-space coarse-graining transformation for the
dimer model on AB graphs, and show that the model has a fixed point under this
transformation. 

\subsection{The coarse-graining transformation}
\label{sec:rgtrans}
Consider a statistical mechanics model with  microscopic
degrees of freedom $s_i$. A configuration $\Sc=\{s_i\}$ is associated
with an weight $W(\Sc)$. The probability of the
configuration $\Sc$ is given by  $P(\Sc)=W(\Sc)/Z$,
where $Z=\sum_{\Sc} W(\Sc)$ is the partition function.

\begin{figure*}[t]
  \includegraphics[width=0.28\columnwidth]{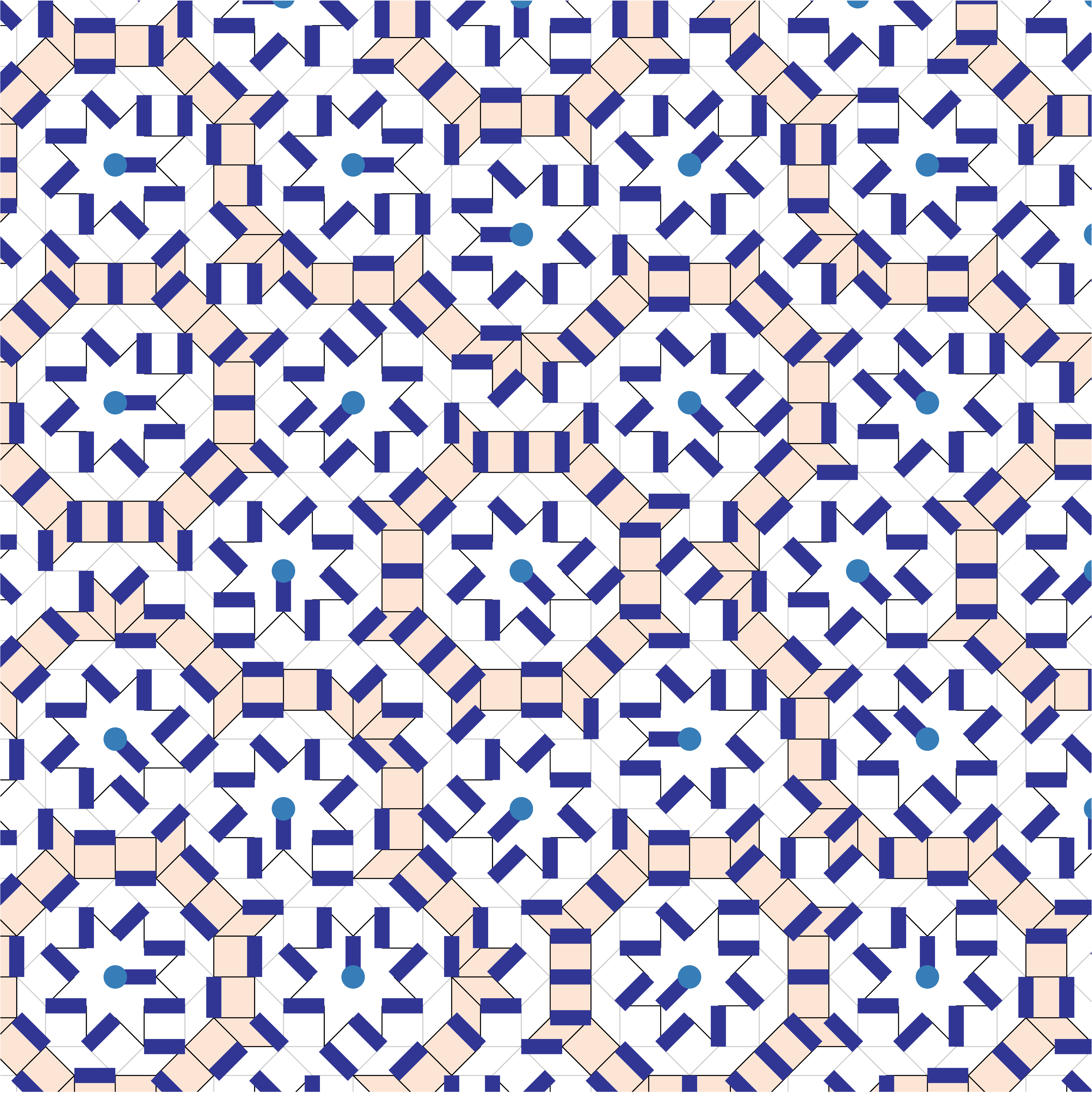}\hspace{1cm}
  \includegraphics[width=0.28\columnwidth]{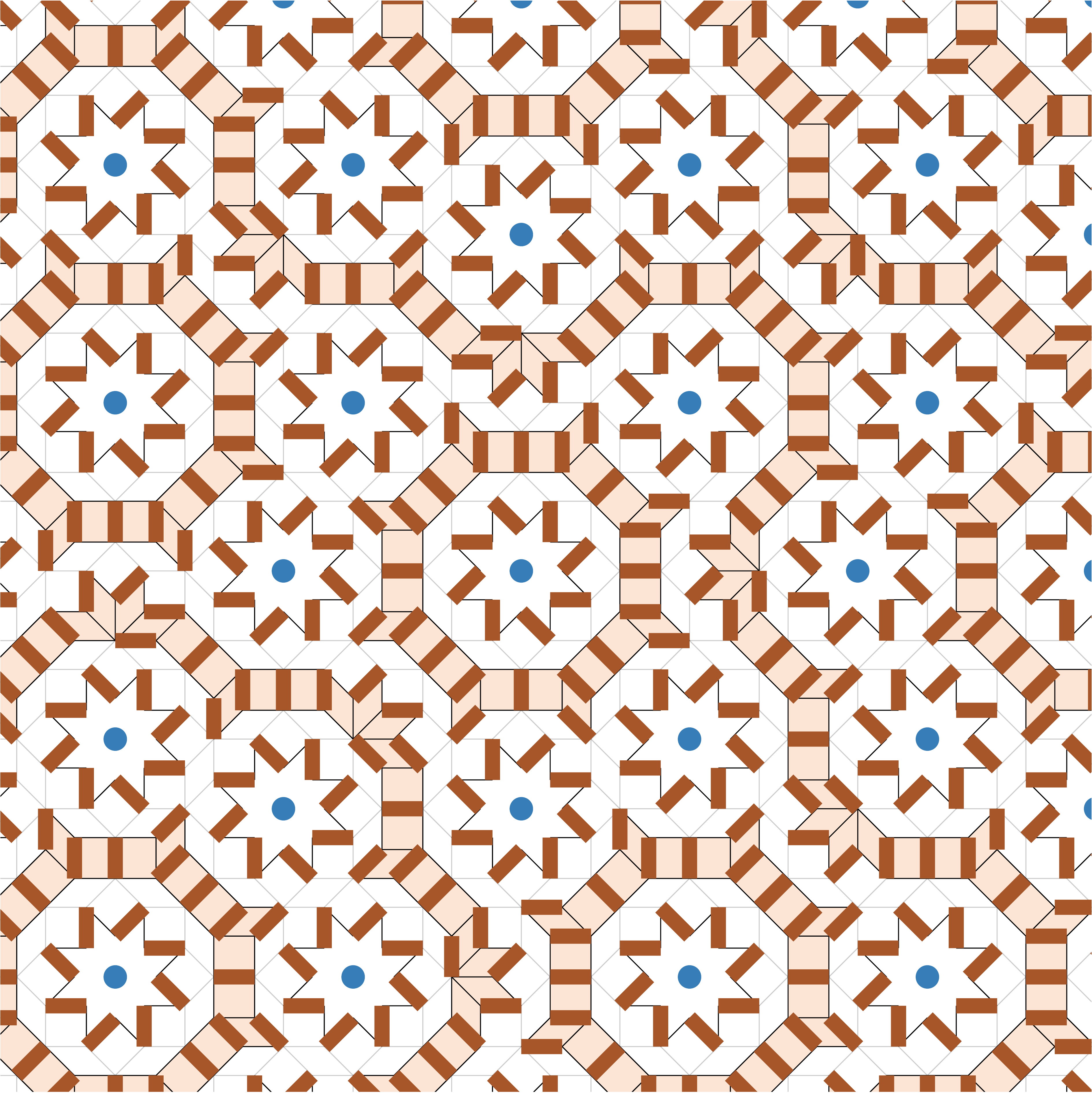}\hspace{1cm}
    \includegraphics[width=0.28\columnwidth]{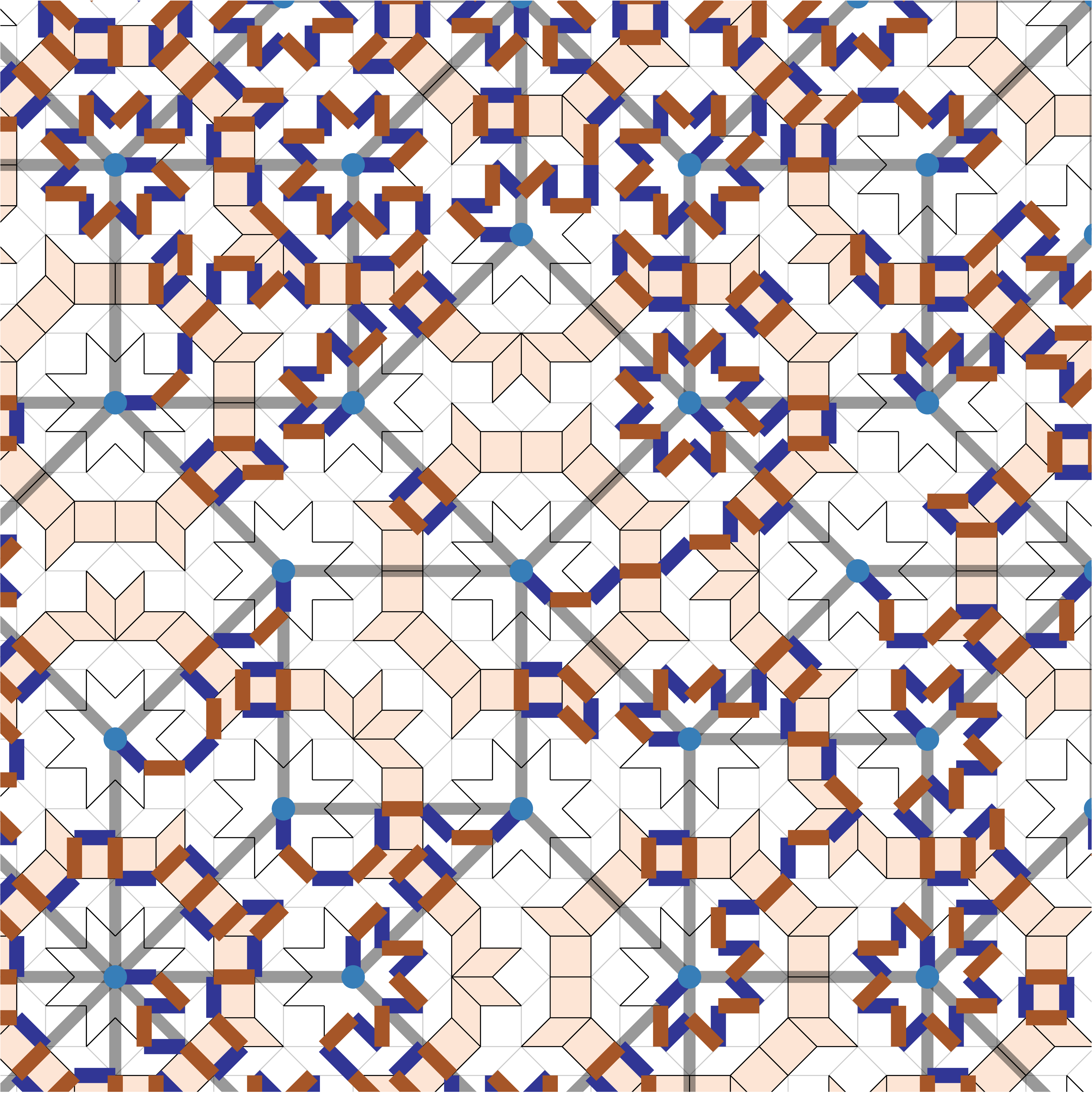}
    \caption{Left: A dimer configuration (blue dimers) on the \AB~graph $\Gc$. Centre: A dimer configuration (brown dimers) on the \ABB~graph $\Gc^{0\backslash 1}$ obtained by removing the $8$-vertices (light-blue solid circles) from $\Gc$. Right: Symmetric difference between brown and blue dimers ---  emptying edges where the blue and brown dimers coincide gives us a configuration of non-overlapping loops and strings with alternating blue and brown dimers. The strings connect the $8$-vertices, which are the vertices of the coarse-grained graph $\Gc^1$; they imply an `effective dimer' of the coarse-grained graph (shown with thick grey lines) in our RG transformation.}
    \label{fig:symmetric_difference}
\end{figure*}
Given a configuration $\Sc$, we introduce a RG transformation $\Lambda$. This is a rule to
generate an ``effective" configuration $\Sc^{\e}$ in terms of effective degrees of
freedom $s^\e_i$, defined in terms of the underlying ``microscopic'' variables $s_i\in \Sc$. More precisely, the transformation $\Lambda$ is defined via  a conditional probability
$P_{\Lambda}(\Sc^\e|\Sc)$, with $\sum_{\Sc^\e}P_{\Lambda}(\Sc^\e|\Sc)=1$ ensuring that the partition function is preserved~\cite{Cardy_book}: we may then write
\begin{align} 
  \nonumber Z=\sum_{\Sc} W(\Sc) &= \sum_{\Sc}\left(\sum_{\Sc^\e} P_{\Lambda}(\Sc^\e | \Sc)\right) W(\Sc) \\ 
  &=\sum_{\Sc^\e}\left(\sum_{\Sc} P_{\Lambda}(\Sc^\e | \Sc) W(\Sc)\right) \\
  &\equiv\sum_{\Sc^\e} W_{\e}(\Sc^\e).  
  \label{eq:rg_def1}
\end{align} 
which defines the weight of an effective configuration $\Sc^\e$,
\begin{equation} 
  W_\e(\Sc^\e)=\sum_{\Sc} P_{\Lambda}(\Sc^\e|\Sc) W(\Sc).
  \label{eq:rg_def2}
\end{equation}
Intuitively, $P_{\Lambda}(\Sc^\e|\Sc)$ can be viewed as a ``projector'' that is nonzero if and only if the microscopic configuration $\Sc$ contributes to the effective configuration $\Sc^{\text{eff}}$.

We now construct such an RG transformation $\Lambda$ which coarse-grains dimer configurations on the \AB~tiling. The transformation is motivated by the structure of perfect-dimer-covered configurations reviewed in Sec.~\ref{sec:review}.
Before formally defining $\Lambda$, we first present an intuitive rule to construct  coarse-grained dimer configurations from microscopic dimer configurations, motivating the rest of the discussion. We begin by reviewing the construction of perfect-dimer covers. Recall that in constructing such a cover, we  first
match all the non-8-vertices (which make up the corresponding auxiliary
\ABB~graph), and then  match pairs of 8-vertices by augmenting alternating paths between
them. These 8-vertices lie on the vertices of the coarse-grained
graph $\sigma^{-2}(\Gc)$.  The second step of this process, of constructing and augmenting alternating  paths between 8-vertices,  resembles that of constructing a dimer cover of the coarse-grained $\sigma^{-2}(\Gc)$ (which consists of {\it only}  the $8$-vertices of $\Gc$).
This suggests that the effective  dimers on the coarse-grained graph $\sigma^{-2}(\Gc)$ should somehow correspond to alternating paths between $8$-vertices on $\Gc$. However, this implies that all the  several alternating paths  between two fixed $8$-vertices in $\Gc$ will map to the same effective dimer on $\sigma^{-2}(\Gc)$. It therefore follows that the weights of the latter should be 
 set entropically by the whole  {\it ensemble} of the former. Constructing this entropic weight represents the central challenge in specifying the RG transformation $\Lambda$.

Starting with an \AB~graph $\Gc$ with dimer configuration $\Sc$, our goal is therefore
to write down an effective dimer configuration $\Sc^{\e}$ on the coarse-grained graph
$\sigma^{-2}(\Gc)$ by ``identifying'' alternating paths between $8$-vertices on $\Gc$ which would have been augmented to obtain the microscopic configuration $\Sc$. 

To accomplish the goal of extracting the alternating 
paths between 8-vertices on $\Gc$ for a given $\Sc$, we overlap the dimers in $\Sc$ with the dimers in a perfect matching
$\Sc^{*}$ of the \ABB~graph (i.e., the graph obtained from the \AB~graph $\Gc$ by removing all 8-vertices). In such overlap
configurations, all 
non-8-vertices have two dimers (one each from $\Sc$ and $\Sc^{*}$), while 8-vertices have a single dimer (from $\Sc$): in other words, 
pairs of $8$-vertices lie on the ends of chains of alternating dimers from $\Sc$ and $\Sc^{*}$, defining an effective dimer between them. The effective dimer configurations constructed by this procedure will evidently
depend  on the auxiliary dimer configuration $\Sc^{*}$; to remove this dependence we will average over $\Sc^{*}$ when defining the RG transformation. It is also sometimes desirable to perform ``larger" RG transformations, \textit{i.e.}, directly write down 
effective dimer configurations on the graph $\sigma^{-2n}(\Gc)$; effective dimers here would correspond to alternating paths between $8_k$-vertices with $k\geq2n$, obtained by overlapping dimer-covers on an $\AB$ graph  with dimer-covers on an auxiliary graph with all $8_k$ vertices for $k\geq2n$ removed. Since such overlaps lead to exactly one alternating path connecting each 8-vertex to another, and the effective dimers so obtained, by construction, obey the hardcore constraint.

To place this intuitive discussion on firmer footing, let us first establish some notation. We will use $\Gc_n$ as a shorthand for the graph $\sigma^{-2n}(\Gc)$, with edge lengths $\delta^{2n}$, that hosts effective dimers obtained after $n$ steps of RG. Vertices of $\Gc_n$ correspond to $8_k$-vertices of $\Gc_0$ with $k \geq 2n$. 
We use the binary variable $s^n_i$ to denote dimer occupancy in the edge $e_i$ of the graph $\Gc_n$, and the set $\Sc^n=\{s^n_i\}$ denotes a dimer configuration on the graph $\Gc_n$.
 We also use the notation
$\Gc_{m\backslash n}$, for $n>m$, to denote the graph obtained by removing all vertices from $\Gc_m$ which also belong to $\Gc_n$, \textit{i.e.,} removing the $8_k$-vertices for $k \geq 2n$. In this language, the \ABB~graph introduced in Sec.~\ref{sec:review} obtained by removing $8$-vertices from the \AB~graph is therefore the graph $\Gc_{0\backslash 1}$.

Adopting this notation, consider a dimer configuration $\Sc=\{s^0_i\}$ constituting a perfect
matching in an \AB~graph $\Gc$, and another dimer configuration $\Sc^{0\backslash 1}=\{s^{0\backslash 1}_i\}$
constituting a perfect matching in the corresponding \ABB~graph $\Gc_{0\backslash 1}$. We consider the
``transposition" (borrowing the terminology from the literature on valence bond wave functions~\cite{BeachSandvik,SandvikEvertz}) of these sets of dimers $\mathcal{T}$, given by
\begin{equation}
\mathcal{T}=\{t_i| t_i\equiv (s^0_i+s^{0\backslash 1}_i)~(\text{mod}~2)\},
\end{equation}
 where $t_i$ is 1 iff either $s^0_i$ or $s^{0\backslash 1}_i$, but not both,
are 1. This corresponds to superposing the dimers from both $\Sc$ and
$\Sc^{0\backslash 1}$ on the same graph and then emptying edges with two dimers. As outlined earlier, the set $\mathcal{T}$ decomposes into different kinds of non-overlapping
components: there are closed loops of alternating dimers  from the sets
$\Sc$ and $\Sc^{1\backslash 0}$; further, there are also open chains of such
alternating dimers which connect the $8$-vertices (lying on the coarse-grained
graph). In our proposed RG transformation such an open chain between two $8$-vertices
imply an effective dimer between them. Fig.~\ref{fig:symmetric_difference}, shows how  transposition between dimer configurations on $\Gc_0$ and $\Gc_{0/1}$ can be used to construct an effective dimer configuration. 

The effective dimer configuration $\Sc^{\text{eff}} \equiv \Sc^1$ will depend not only on the
microscopic dimer configuration $\Sc^0$ but also the auxiliary configuration
$\mathcal{S}^{0\backslash 1}$ on the corresponding \ABB~graph; this dependence is taken into account by
defining the coarse-graining  rule $\Lambda$ as an average over all such
auxiliary configurations $\mathcal{S}^{0\backslash 1}$. Formally, we write
\begin{align}
\nonumber  Z= \sum_{\Sc^0}W(\Sc^0)= \sum_{\Sc^1,\Sc^0} P_{\Lambda} (\Sc^1 | \Sc^0)W(\Sc^0), \text{with} \\
  P_{\Lambda}(\Sc^1|\Sc^0) = \sum_{\Sc^{0\backslash 1}} \vartheta_{\Sc^1}(\Sc^0,\Sc^{0\backslash 1}) P(\Sc^{0\backslash 1}).
  \label{eq:rg_transf1}
\end{align}
The function  $\vartheta_{\Sc^1}(\Sc^0,\Sc^{0\backslash 1})$ equals 1 if the transposition
graph of $\Sc^0$ and $\Sc^{0\backslash 1}$ lead to the effective dimer configuration $\Sc^1$, 0
otherwise. $P(\Sc^{0\backslash 1})$ is the uniform probability distribution over perfect dimer covers of the
auxiliary \ABB~graph $\Gc_{0\backslash 1}$. This leads to effective dimer configurations on the 
graph $\Gc_1$. The effective dimer partition function can be seen as being defined by Eq.~\eqref{eq:rg_transf1}  to be an average over a ``double ensemble"
of perfect matchings: one over the \AB~graph $\Gc_0$, and the other over the auxiliary ~\ABB~graph $\Gc_{0\backslash 1}$. 

The generalisation to ``larger'' RG transformations is now immediate. We simply obtain a transposition between perfect matchings on the graphs $\Gc_0$ and 
$\Gc_{0\backslash n}$, to obtain effective dimer configurations on the graph $\Gc_n$ with edge lengths $\delta_s^{2n}$. The average over auxiliary configurations $\Sc^{0\backslash n}$ defines the $n$-step RG transformation
\begin{align}
  P_{\Lambda}(\Sc^n|\Sc^0) = \sum_{\Sc^{0\backslash n}} \vartheta_{\Sc^n}(\Sc^0,\Sc^{0\backslash n}) P(\Sc^{0\backslash n}),
  \label{eq:rg_transf2}
\end{align}
where $P(\Sc^{0\backslash n})$ denotes the uniform probability distribution over perfect dimer covers on the auxiliary graph $\Gc_{0\backslash n}$.

One might expect that the large RG transformation described above can also by 
implemented by composing two smaller RG transformations, \textit{i.e.}, by first obtaining effective configurations at an intermediate scale $\Sc^m$ for some $0<m<n$, and then using that to obtain $\Sc^n$. We compose
the two transformations $P_\Lambda(\Sc^m|\Sc^0)$ and $P_\Lambda(\Sc^n|\Sc^m)$ as follows:
\begin{align}
  \nonumber P_{\Lambda} (\Sc^m|\Sc^0) &= \sum_{\Sc^{0\backslash m}} \vartheta_{\Sc^m}(\Sc^0,\Sc^{0\backslash m}) P(\Sc^{0\backslash m}), \\
  \nonumber P_{\Lambda} (\Sc^n|\Sc^m) &= \sum_{\Sc^{m\backslash n}} \vartheta_{\Sc^n}(\Sc^m,\Sc^{m\backslash n}) P(\Sc^{m\backslash n}),\text{ and}\\
  \overline{P_\Lambda}(\Sc^n|\Sc^0) &= \sum_{\Sc^m} P_\Lambda(\Sc^n|\Sc^m) P_\Lambda(\Sc^m|\Sc^0).
  \label{eq:rg_transf_compose}
\end{align}
Note that the distribution $P(\Sc^{m\backslash n})$ over effective dimer configurations on the graph $\Gc^{m\backslash n}$  $P(\Sc^{m \backslash n})$ is {\it not} uniform, but can in principle be   obtained by coarse-graining $P(\Sc^{0\backslash n})$, the distribution over perfect dimer covers on $\Gc^{0\backslash n}$, as follows:
\begin{align}
  P_{\Lambda} (\Sc^{m\backslash n}|\Sc^{0\backslash n}) &= \sum_{\Sc^{0\backslash m}} \vartheta_{\Sc^{m\backslash n}}(\Sc^{0\backslash n},\Sc^{0\backslash m}) P(\Sc^{0\backslash m}). 
  \label{:eq:rg_transf_neverrefertothis}
\end{align}
It is tempting to 
assume that the composed transformation, $\overline{P_\Lambda}(\Sc^n|\Sc^0)$ of Eq.~\eqref{eq:rg_transf_compose} is equal to the transformation $P_\Lambda(\Sc^n|\Sc^0)$ from Eq.~\eqref{eq:rg_transf2}: 
such composed transformations naturally lend themselves to RG iterations, where successive RG transformations lead to 
effective descriptions at larger and larger scales. However, analysis of our numerical implementation of the RG reveals that this is not in fact the case --- and indeed, this scale-composition law is rarely true of similar real-space `block' RG schemes. Nevertheless, the direct and composed transformations are equal ``in the RG sense'': the effective Hamiltonians they generate only differ by irrelevant operators~\footnote{Specifically, we have  numerically verified that all sizeable dimer correlations are equal to within statistical error for the two different transformations. The differences are in the occupancies of NNN dimers, which remain very low in both cases and do not affect correlations further ``downstream''  in the RG and are in this sense irrelevant variables.}. Therefore to streamline our discussion we will often ignore the distinction between $P_\Lambda(\Sc^n|\Sc^0)$ and $\overline{P_\Lambda}(\Sc^n|\Sc^0)$.

\subsection{Numerical evidence of the fixed point }
\label{sec:mcrg}
\begin{figure}[t]
    \includegraphics[width=\columnwidth]{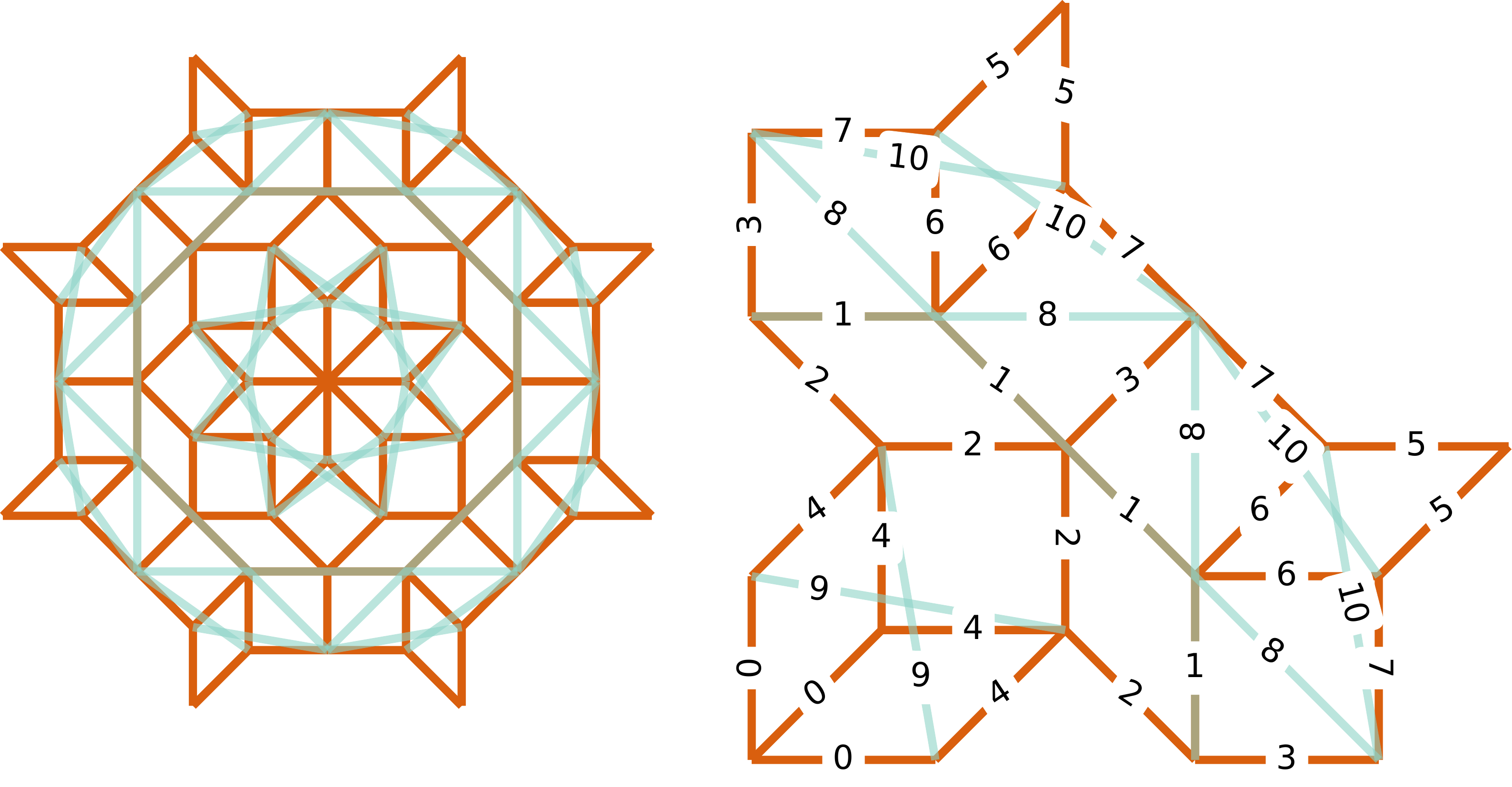}
    \caption{Left: An $8_1$-region, with edges in orange, and NNN-edges in light blue. While the microscopic model does not have any dimers on NNN edges, our RG transformation generates dimers on the NNN edges. Right: The labelling scheme to uniquely identify all (including NNN) symmetry inequivalent edges of the $8_1$ region. We use these labels to report dimer densities in Tab.~\ref{tab:dim-den} as we track them as the Hamiltonian flows to a fixed point.}
    \label{fig:8_1_region}
\end{figure}
We now study the effective dimer configurations generated
by the transformations presented in Eqs.~\eqref{eq:rg_transf1} and
~\eqref{eq:rg_transf2} from the preceding section.  Our main workhorse for this section (and the rest of
the paper) is Monte Carlo Renormalisation Group (MCRG)~\cite{Swendsen_PRL}. The basic idea
involves sampling microscopic configurations $\Sc$ using standard Monte Carlo
techniques, and then coarse-graining each configuration $\Sc$ to an effective
configuration $\Sc^{\e}$, thereby generating samples of effective configurations
$\Sc^\e$. This allows the calculation of observables in terms of effective
degrees of freedom. In usual critical phenomena, correlations between microscopic and coarse-grained degrees of freedom also allow the estimation of critical exponents. To implement MCRG for our dimer model, we use the standard
directed loop algorithm for classical dimer models~\cite{Sandvik_Moessner,Alet_etal}.
We generate samples of effective configurations $\Sc^n$ on the graph $\Gc^n$ by coarse graining the configurations $\Sc^0$  on graph $\Gc^0$ using the rule $\Lambda$ described in Sec.~\ref{sec:rgtrans}. To do so, we first generate 
independent samples from two different ensembles: the ensemble of perfect matchings on the graph $\Gc^0$, and the ensemble of perfect matchings on the graph $\Gc^{0\backslash n}$(obtained by removing from $\Gc^0$ all vertices which belong to $\Gc^n$). Given these two dimer configurations, we transpose them to obtain effective configurations $\Sc^n$ on the graph $\Gc^n$, as described in Eq.~\eqref{eq:rg_transf2} and illustrated in Fig.~\ref{fig:starsandladders}.

Due to the lack of translational invariance and the intrinsic impossibility of imposing periodic boundary conditions, investigations on quasiperiodic graphs such as \AB~ must work with
specific samples and boundary conditions. For some purposes, it is convenient to work with periodic
approximants to these graphs. Here, instead, given the focus of our investigations  of the dimer problem, we wish to work with samples which preserve
the discrete scale symmetry of the graph and its $D_8$-symmetries to the maximum extent
possible given the finite-size testriction.  We choose the $8_n$ regions, or effective vertices, introduced in Sec.~\ref{sec:review} (cf.  Fig.~\ref{fig:effective_vertices}) 
as the finite graphs for our numerical investigations. This choice offers two distinct advantages apart from the explicit $D_8$-symmetry. Firstly, matching problems in any large patch of an \AB~graph have
descriptions in terms of effective matching problems between $8_n$ regions.
$8_n$ regions are ``effective vertices'', \textit{i.e.}, they collectively have exactly one dimer
between them and the vertices in the rest of the graph; intuitively, the
matching problem on any large graph coarse-grains to an effective matching
problem of an $O(1)$ number of these $8_n$ regions.  Secondly, 
each $8_n$ region has exactly
one monomer in its maximum matching. This ensures that the choice of boundary
conditions do not introduce an unnecessarily large number of  monomers whose
effects, while expected to be irrelevant in the thermodynamic limit, might significantly
modify results on finite-size samples. 

$8_n$ regions are inflations of each other; coarse-graining a dimer
configuration on an $8_{2n+1}$ region using the $n$-step RG transformation $\Lambda^n$
(Eq.~\eqref{eq:rg_transf2}) leads to an effective dimer configuration on an
$8_1$-region.  We  compare the effective dimer observables on the $8_1$-regions
obtained from coarse-graining $8_{2n+1}$ regions with the coarse-graining rule
$\Lambda^n$ for different $n$, and see that they reach a fixed point with
increasing $n$.  We consider $n=(0,1,2,3)$; the corresponding
$8_{2n+1}$-regions are graphs with 48, 2497, 87425 and  2993281 vertices
respectively. 

\begin{figure}[t]
    \includegraphics[width=6cm]{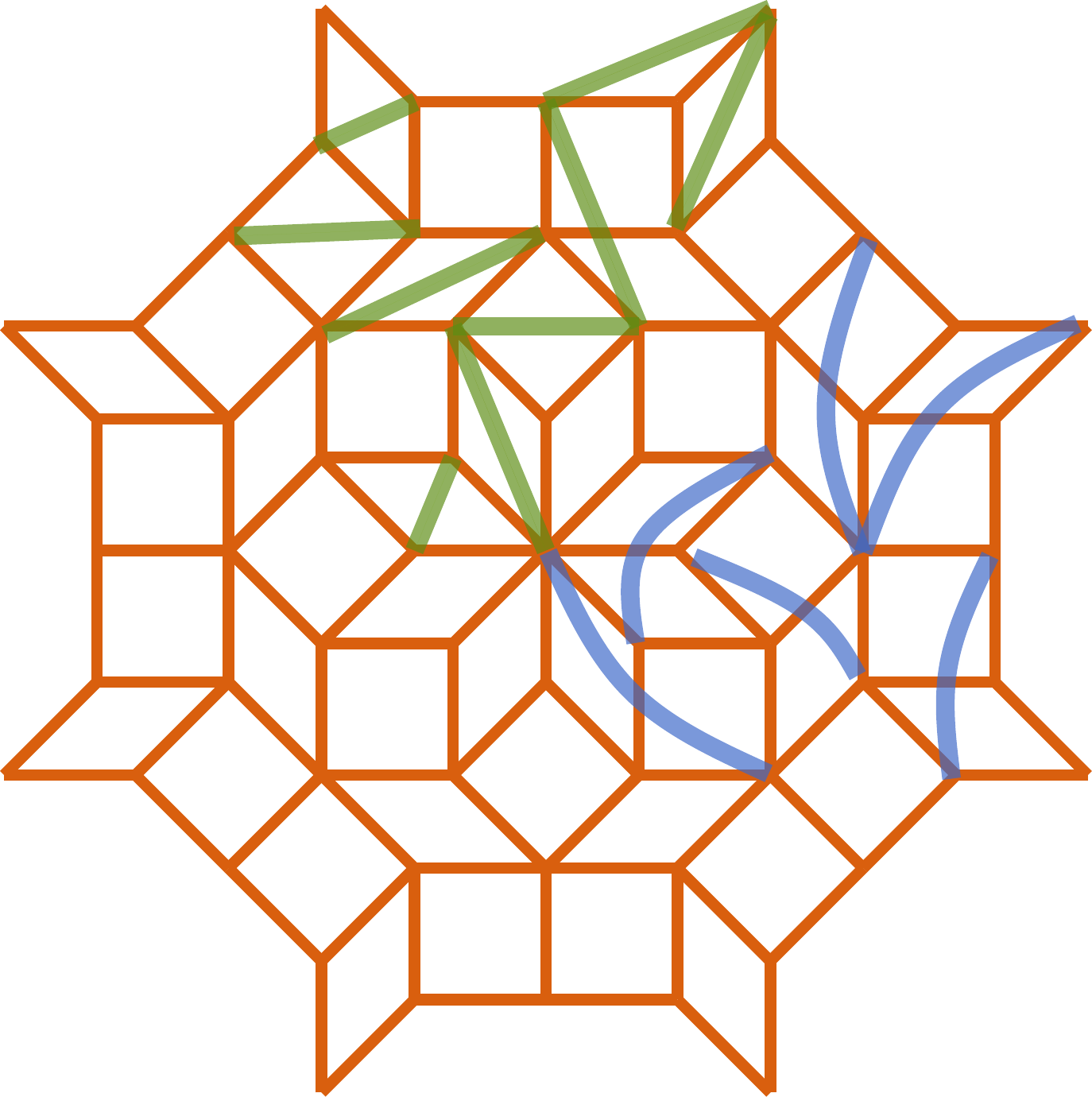}
    \caption{ Our RG transformation generates effective dimers between vertices of opposite sublattice; the most important  new contributions are from next-next-neighbour (NNN) vertices, \textit{i.e.,} ones separated by three edges. Some representative NNN vertex pairs which can host an effective dimer are marked with blue lines  for an $8_1$-region. Note that our RG transformation does not generate effective dimers between  vertices of the same sublattice; therefore, next-neighbour (NN) vertex pairs (marked with green lines) separated by two edges never host an effective dimer.}
    \label{fig:nnn_dimers}
\end{figure}

In Fig.~\ref{fig:8_1_region}, we display the $8_1$-region on which we compare
effective dimer models obtained by coarse-graining microscopic dimer models on
different $8_{2n+1}$-regions. We also assign unique labels to all the
symmetry-inequivalent edges to set up the notation. Note, that the RG transformation ``generates" dimers between vertices that are separated by three edges; vertices separated by two edge-distances belong to the same sublattice,  and dimers between such vertices can never be generated by our RG transformation. We will call these dimers between vertices separated by three edges  next-next-neighbour(NNN)-dimers (see Fig.~\ref{fig:nnn_dimers}).

We first look at the effective dimer densities. The dimer density on an edge
$e_i$ is defined as the fraction of dimer configurations that host a dimer on
the edge $e_i$. We see that for all edges, the dimer densities reach a fixed
point density within $n=3$ RG steps. These dimer densities, and their
convergence to a fixed point, are shown in Tab.~\ref{tab:dim-den}. We also note
that the densities of generated NNN dimers remain quite low ($\sim 10^{-3}$) as the fixed point is approached.

While this provides strong evidence that the effective dimer densities are at a fixed point, we expect that the fixed point also forces the same fate for the entire dimer probability
distribution, as probed by different correlation functions.  To explore this, we also compute connected correlation functions of effective
dimers; the connected dimer correlation between two dimers on edges $e_i$ and
$e_j$ is given by 
\begin{equation}
  C^n(e_i,e_j) = \langle s^n_i s^n_j \rangle -\langle s^n_i \rangle \langle s^n_j \rangle.
\end{equation}

We find that dimer correlations also flow to a fixed point within a few RG
iterations; for brevity, we  tabulate the numerical
values of dimer correlations showing their convergence to the fixed point in
Appendix~\ref{app:dimercorrs}.

We remark that the choice of $8_1$-region, to which we coarse-grain dimer configurations on
$8_{2n+1}$-regions in this analysis, is driven primarily by the  numerical advantages offered by its
relatively smaller size and is not an essential property of the RG.  We have verified that starting with dimer
models on $8_{2n+2}$-regions and coarse-graining them to effective dimer models
on $8_2$-regions also leads to a fixed point of effective dimer densities with
increasing $n$.

\begin{table}[!t]
  \centering
  \begin{tabular}{ |p{0.11\columnwidth}| p{0.16\columnwidth}| p{0.16\columnwidth} | p{0.16\columnwidth}| p{0.16\columnwidth}|}
  \hline
 Edge&\multicolumn{4}{|c|}{Effective-dimer density after $n$ RG steps} \\
\hline
    & n=0 & n=1 & n=2 & n=3 \\
\hline
\hline
0 & 0.1244 & 0.1243 & 0.1242 & 0.1243 \\
\hline
1 & 0.2074 & 0.1942 & 0.1921 & 0.1922 \\
\hline
2 & 0.0521 & 0.0458 & 0.0456 & 0.0455 \\
\hline
3 & 0.4811 & 0.5188 & 0.5237 & 0.5235(1) \\
\hline
4 & 0.4378 & 0.4374 & 0.4375 & 0.4375(1) \\
\hline
5 & 0.4798 & 0.4915 & 0.4916 & 0.4917(1) \\
\hline
6 & 0.2809 & 0.2897 & 0.2918 & 0.2916(1) \\
\hline
7 & 0.2393 & 0.2125 & 0.2115 & 0.2116(1) \\
\hline
8 & 0.0000 & 0.0007 & 0.0005 & 0.0005 \\
\hline
9 & 0.0000 & 0.0005 & 0.0004 & 0.0004 \\
\hline
10 & 0.0000 & 0.0063 & 0.0051 & 0.0051 \\
\hline
\end{tabular}
\caption{Starting with the graph $\Gc$ being an $8_1$ region of Fig.~\ref{fig:8_1_region}, we consider a series of inflated graphs $\Gc_n=\sigma^{2n}$ for $n=0,1,2,3$. We use our RG transformation $n$ times for the graph $\Gc_n$ to coarse grain it back to the $8_1$-region $\Gc_0$ and compute effective dimer densities there. We use the edge-labelling for symmetry-inequivalent edges introduced in Fig.~\ref{fig:8_1_region} to report the dimer densities, and we see that they reach a fixed point with increasing $n$, implying that the coarse-grained dimer problem on the $8_1$-region has approached a fixed point. In Tab.~\ref{tab:dim-corr}, reported in Appendix~\ref{app:dimercorrs}, we report dimer correlations and see that they also approach a fixed point with increasing $n$.  }
\label{tab:dim-den}
\end{table}

\section{The effective Hamiltonian}
\label{sec:effham}
Having provided strong evidence suggesting that the distribution of effective
dimers converges to a fixed point under the RG transformation defined by
Eq.~\eqref{eq:rg_transf2}, we now proceed to calculate the Hamiltonians which
describe the distribution of effective dimers. An expression for a Hamiltonian governing effective dimers provides a direct characterisation of the fixed point. Additionally, this yields another advantage:  an expression for such effective Hamiltonians would allow us to sample effective dimer distributions on a graph directly, opening up the possibility to use the RG  to effectively probe  larger system sizes. 

\subsection{Strategy to calculate effective Hamiltonian}
\label{sec:effham_method}

The microscopic dimer model is
purely entropic and ascribes no energy cost to any allowed configuration, i.e.  each dimer configuration has exactly the same weight. On the other hand, the definition
of our RG transformation (Eqs.~\eqref{eq:rg_transf1} and ~\eqref{eq:rg_transf2}) already points to the reason
why effective dimer configurations generated by it might acquire weights
relative to each other. In our transformation, effective dimers correspond to open paths between
$8$-vertices obtained by transposing two different dimer covers: a perfect
dimer cover on the \AB~tiling, and one on the auxiliary \ABB-tiling (which
does not have 8-vertices).  The relative weights of effective dimer configurations are then
set by the total number of such open paths for each effective dimer
configuration, counted in the double ensemble defined by  perfect matchings on both
the \AB~ and \ABB~graphs. Evidently, there is no {\it a priori} reason for these weights to be identical given the lack of translational invariance. However, calculating such weights in full generality is a formidable challenge.

The difficulty can be alleviated by the assumption, to be justified later, of ``weight factoring'': namely, that
the weight function of an effective dimer configuration is well approximated by
the product of weights of  dimers at each edge, or equivalently, that the interactions
between effective dimers are negligible. For effective dimer configurations $\Sc^n=\{s^n_i\}$, if $W_n(e_k)$ denotes the weight of a dimer on edge $e_k$, this amounts to assuming that the partition function of the weighted dimer model can be written as
\begin{align}
  Z \sim \sum_{\Sc^n} \prod_k (W_n(e_k))^{s^n_k}.
  \label{eq:partition_func}
 \end{align}
With this assumption, the problem of calculating $W_n(e_k)$, the weight
of occupation of a specific edge by an effective dimer, can be posed in a
tractable manner, as follows.

Consider, in full generality, an effective configuration $\{s^{n}_i\}$ on the graph $\Gc_{n}$ . We first calculate the effective dimer
density $\langle s^n_k \rangle$  on the edge $e_k$, which, under our assumption of decoupled dimer weights, reduces  to
\begin{align}
  \langle s^n_k \rangle = W_n(e_k) \sum_{\Sc^{n}/e_k} \prod_{m \neq k} (W_n(e_m))^{s^n_m}.
  \label{eq:dimerden_eff}
\end{align}
The sum over $\Sc^n/e_k$ is a sum over all dimer configurations with the condition that there is always a dimer on $e_k$. The weight $W_n(e_k)$ has the expression:
\begin{align}
  W_n(e_k)= \frac{\langle s^n_k \rangle}{ \sum_{\Sc^{\e}/e_k} \prod_{m \neq k} (W_n(e_m))^{s^n_m}}.
  \label{eq:wt_calc}
\end{align}
The numerator of this expression is directly accessible to Monte Carlo simulations; we calculate
the expectation value of an effective dimer by coarse-graining microscopic
dimers as described in Sec.~\ref{sec:rgtrans}. From Eq.~\eqref{eq:dimerden_eff},
it is clear that the denominator is the dimer density $s^\e_k$ in a different
dimer model with effective dimer weights $W_\e(e_m)$ on all edges $e_m$ for $m
\neq k$, and  weight 1 for an effective dimer on $e_k$. This observation allows us to devise a  Monte-Carlo algorithm which 
calculates the weights $W_n(e_k)$  in Eq.~\ref{eq:wt_calc} for all edges $e_k$ in
one go, starting from dimer configurations in the microscopic graph $\Gc_m$ for $m<n$ . This algorithm to calculate the weights $W_n(e_k)$ overcomes a key technical challenge, allowing us to calculate and investigate effective Hamiltonians directly. In the interest of maintaining the linearity of presentation, we defer a detailed presentation of the algorithm and its subtleties to Appendix~\ref{app:algorithm}, and move directly to a discussion of the effective Hamiltonian.

\begin{figure}[t]
    \includegraphics[width=\columnwidth]{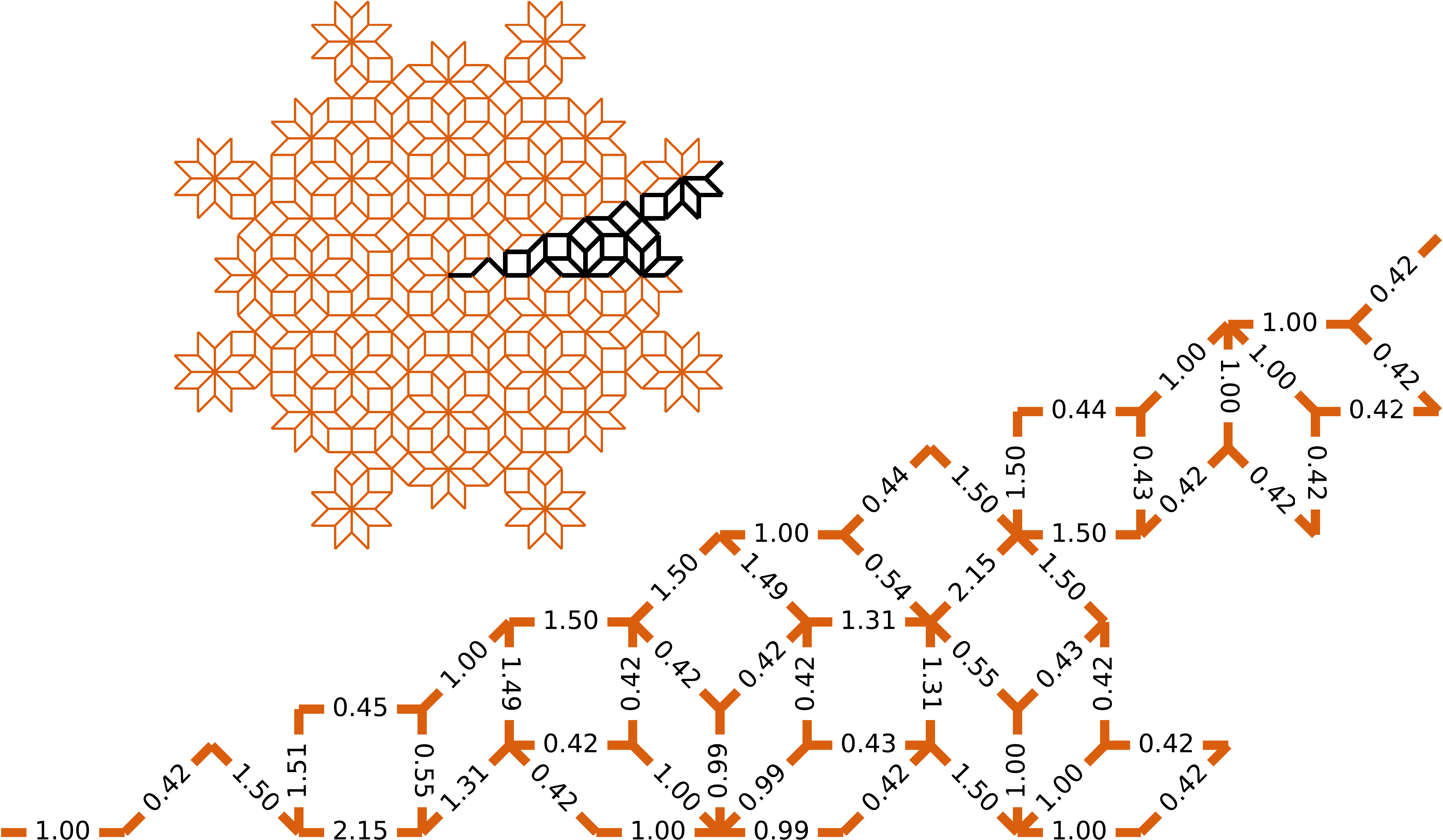}
    \caption{ Left: an $8_2$-region. Right: The effective Hamiltonian  of an $8_2$-region, given in terms of dimer weights of its symmetry-inequivalent edges.  These are obtained using the methods of Sec.~\ref{sec:effham_method}, by coarse-graining the microscopic dimer problem (with all dimer weights 1) on an $8_4$-region.}
    \label{fig:dimer_weights}
\end{figure}

\subsection{Structure of the effective Hamiltonian}
\label{sec:effham_results}

\begin{figure}[t]
    \includegraphics[width=\columnwidth]{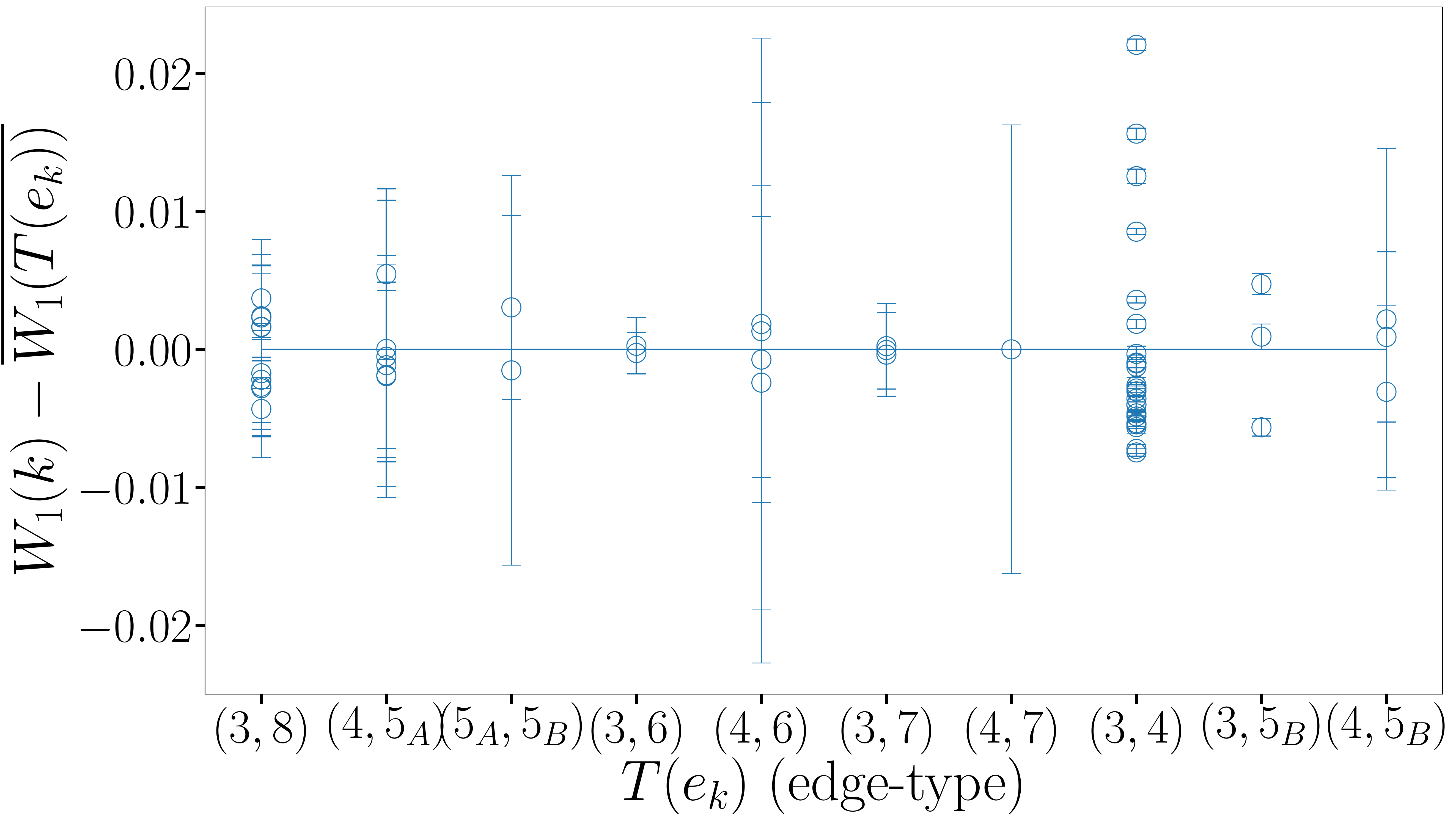}
    \caption{We calculate the effective dimer weights $W_1(k)$ obtained by coarse-graining the microscopic dimer problem once   on an $8_4$-region to an effective problem on an $8_2$ region. Here, we show the difference of $W_1(k)$ from the edge-type average $W_1(T(e_k)$, sorted by the edge-type $T(e_k)$ on the $x$-axis. The figure shows that the differences are small, indicating that the dimer weight obtained under RG for an edge is set primarily by the edge-type. We ascribe the small differences to the presence of dimer-interactions.}
    \label{fig:edge_weights_type}
\end{figure}

We now use the algorithm described above to calculate the weights of dimers
$W_n(e_k)$ after $n$ rescalings. 
We begin with the microscopic dimer problem on an $8_4$
region and use the algorithm of Appendix~\ref{app:algorithm} to calculate weights $W_1(e_k)$ of dimers on the coarse-grained $8_2$ region.
We display the $8_2$ region, and the calculated weights $W_1(e_k)$ for a set of
its symmetry inequivalent edges in Fig.~\ref{fig:dimer_weights}. 
These 
weights specify the effective Hamiltonian $\Hc_1$ after one RG step on the $8_2$ graph, and hence the partition
function is given by Eq.~\eqref{eq:partition_func}. To illuminate the structure
of this effective Hamiltonian, we note that the weight of an edge depends very strongly on the ``type" of the edge. The type $T(e_k)$ of an edge $e_k$ is determined by the  
the vertex types to which the edge connects: recall that there are $7$ types of vertices,  introduced in Sec.~\ref{sec:review}: $3$, $4$, $5_A$, $5_B$, $6$, $7$ and $8$. For example, if an edge $e_k$ connects an $8$-vertex to a $3$-vertex, the edge type $T(e_k)=(3,8)$. To check this quantitatively we calculate $\overline{W_1(t)}$, the average weight of all edges with type $T(e_k)=t$.  In Fig.~\ref{fig:edge_weights_type}, we plot 
the difference of edge weight and the type average $W_1(e_k) -\overline{W_1(T(e_k))}$. We see that the differences are small,
and typically are $\lesssim 2\%$ of the edge weights themselves.

To see intuitively why this strong edge-type dependence emerges,  recall that (as described in Sec.~\ref{sec:effham_method}) the weights of
effective dimers in $\Gc_{n+1}$ measure the number of open alternating paths between
8-vertices of $\Gc_n$ obtained in the double ensemble of two matchings: one on $\Gc_{n}$ and another on $\Gc_{n\backslash n+1}$.
It is reasonable to expect that  these weights will depend on the structure of
$\Gc_n$ in the neighbourhood of the pair of 8-vertices in question. Since the local neighbourhood of each of these 8-vertices are determined by their vertex-type in $\Gc_{n+1}$(Sec.~\ref{sec:review}), it follows that the weight of an effective dimer on an edge $e_k$ will depend strongly on the type of the two vertices connected by $e_k$. 

This motivates us to propose an approximation $\overline{\Hc_n}$  to the effective Hamiltonian $\Hc_n$ after $n$ RG steps: $\overline{\Hc_n}$ involves using the edge-type average weight $\overline{W_n(t)}$ for all edges $e_k$ with an edge-type $T(e_k)=t$, i.e. it asserts that the edge weights are functions solely of the edge type.
The small deviations of the calculated weights $W_1(k)$ from the edge-type average
$\overline{W_1}(T(e_k))$ are then presumably the result of weak interactions between
effective dimers, arising from the constraint that the open alternating paths in the double
ensemble that correspond to effective dimers cannot intersect.  To wit, if two effective edges share a vertex,
the only interactions between them arise out of the hardcore constraint, but two parallel edges that share a plaquette, 
might be associated with interactions arising out of the non-intersection constraints of the underlying alternating paths.  We can incorporate this effect with $\oo{\Hc}$,
a better approximation to the effective Hamiltonian. In $\oo{\Hc_n}$, we allow the weight of an effective dimer on an edge $T(e_k)$ to depend not only on the edge-type $T(e_k)$, but also the types of the two edges parallel to $e_k$. Thus, $\oo{\Hc_n}$ is parameterised by the weights $\oo{W_n}(t,t^{||}_1,t^{||}_2)$. 

Note that both types of effective Hamiltonians lead to identical interactions for different edges that have similar local environments, and as such attempt to capture the notion of `translational invariance' as closely as possible in the quasiperiodic environment. These approximations leads to significant simplifications.  Specifying $\overline{\Hc_n}$ requires just 10 parameters for graphs of all sizes, viz. a dimer weight for each of the 10 edge types, which is a considerable reduction from specifiying independent weights for each symmetry-inequivalent edge. Similarly, $\oo{\Hc_n}$ requires a larger number of parameters but still $O(1)$ as opposed to $O(N)$.
Specifying the effective Hamiltonian with $O(1)$ parameters is crucial to perform the RG iteratively, and allows us to track the effective Hamiltonian as the RG proceeds. 

To do this, we have coarse-grained the microscopic dimer problem on an $8_4$-region to an $8_2$ region to calculate the dimer weights $W_1(e_k)$ after one step of RG for all edges of the $8_2$-region. Now, using the approximation $\oo{\Hc_1}$, \textit{i.e.}, the average weights $\oo{W_1}$, we can sample 
the effective Hamiltonian $\oo{\Hc_1}$ directly on an $8_4$ graph. Coarse-graining 
this to an $8_2$ graph again, we calculate the weights $W_2(e_k)$ after two steps of RG. We can now use $W_2(e_k)$ to construct the approximate effective Hamiltonian $\oo{\Hc_2}$ given in terms of weights $\oo{W_2(t})$. Coarse-graining $\oo{\Hc_2}$ on an $8_4$-graph now give us the weights $W_3(e_k)$ after 3 RG steps.  Iterating this procedure allows us to track the RG flow of the effective Hamiltonian $\Hc_n$, in terms of the dimer weights $\oo{W_n}$, with increasing RG step $n$. In this process, we find that the weights
of dimers maintain a strong dependence on the edge-type for all RG steps, \textit{i.e.}, for all $n$, the weights $W_n(e_k)$ are narrowly distributed around the edge-type average $\overline{W_n}(T(e_k))$ (as displayed in Fig.~\ref{fig:edge_weights_type} for the first RG step). To track the effective Hamiltonian $\Hc_n$, we display the 
weights $\overline{W_n}(t)$ for all edge-types as a function of the RG step $n$ in Tab.~\ref{tab:effham_flow}.
It is evident that our effective Hamiltonians flow
to a fixed point Hamiltonian. Thus, we have not only provided strong evidence for a fixed-point in the dimer problem on the \AB~graph, but also calculated a simple and explicit expression for the fixed-point Hamiltonian given in terms of weights of dimers.

We close this section by mentioning, that in Appendix~\ref{app:effham_details}, we have shown that the effective Hamiltonian approximations used here (both $\overline{\Hc}$ and $\oo{\Hc}$) in terms of the effective dimer weights provide an accurate description of the distribution of effective dimers. To do this,  we have first calculated effective dimer observables obtained directly from a dimer model with the calculated effective dimer weights, and compared them with effective dimer observables obtained by coarse-graining the microscopic Hamiltonian using MCRG simulations.
Note that this automatically implies that the initial hypothesis that the effective Hamiltonian involves the effective dimers simply acquiring weights in the partition function (with dimer-dimer interactions being negligible) is also
accurate. We have also demonstrated in Appendix~\ref{app:effham_details} that the errors introduced by making the approximations $\overline{\Hc}$ vs $\oo{\Hc}$ are irrelevant in an RG sense, \textit{i.e.}, differences between them  disappear with further RG transformations. Therefore it suffices to consider the simplified effective Hamiltonian $\overline{\Hc}$ when analyzing universal properties.

\begin{table}[t]
\begin{tabular}{|c|c|c|c|c|c|}
  \hline
  Edge-type $t$ &\multicolumn{5}{|c|}{Effective dimer weight $\ovl{W_n(t)}$ after $n$ RG steps} \\
  \hline
   & $n=0$ & $n=1$ & $n=2$ & $n=3$ & $n=4$ \\
\hline
\hline
$(3,4)$	& 1.000	& 0.426(1)	& 0.486(1)	& 0.485(1)	& 0.485(1)	\\ 
\hline 
$(3,6)$	& 1.000	& 1.002	& 1.003	& 1.001	& 1.002	\\ 
\hline 
$(3,7)$	& 1.000	& 1.003	& 0.999	& 1.001	& 1.001	\\ 
\hline 
$(3,8)$	& 1.000	& 1.000	& 1.000	& 1.000	& 1.000(1)	\\ 
\hline 
$(3,5_B)$	& 1.000	& 0.547(2)	& 0.590(2)	& 0.589(2)	& 0.588(1)	\\ 
\hline 
$(4,5_A)$	& 1.000	& 1.505(1)	& 1.461(1)	& 1.456(1)	& 1.459(1)	\\ 
\hline 
$(4,6)$	& 1.000	& 1.499	& 1.456(1)	& 1.455(1)	& 1.455(1)	\\ 
\hline 
$(4,7)$	& 1.000	& 1.501	& 1.450	& 1.458	& 1.456	\\ 
\hline 
$(4,5_B)$	& 1.000	& 1.315(1)	& 1.313(2)	& 1.312(1)	& 1.312(3)	\\ 
\hline 
$(5_A,5_B)$	& 1.000	& 2.157(1)	& 2.021(1)	& 2.022(7)	& 2.020(3)	\\ 
\hline 
\end{tabular}
\caption{ Starting with dimer weights $\ovl{W_0(t)}$ for the microscopic Hamiltonian, at each step we coarse-grain the approximate effective Hamiltonian $\oo{\Hc}$ on an $8_4$-region, given by the dimer weights $\ovl{W_n(t)}$,  to an $8_2$-region, calculating the new effective dimer weights by the methods introduced in Sec.~\ref{sec:effham_method}. We track effective Hamiltonians $\ovl{\Hc}$  under RG in terms of the weights $\ovl{W_n(t)}$; showing that they flow to fixed point values $\ovl{W_n(t)}^{*}$ under our RG transformation.}
\label{tab:effham_flow}
\end{table}

\subsection{Relevance of dimer and monomer fugacities at the fixed point}
The fact that a Hamiltonian parameterised in terms of the weights  $\overline{W_n}(t)$ of effective
dimers on each edge-type flows to the fixed point under a numerical procedure
already indicates that the fixed point is stable both to small perturbations in  the edge-type weights $\overline{W_n}(t)$, i.e. that these perturbations are irrelevant in the RG sense. This is confirmed by a more quantitative calculation, presented in Appendix~\ref{app:dimer_irrelevance}. We expect other perturbations (like weights for next-nearest-neighbour dimers which break the bipartiteness of the graph) to be relevant, but calculating  scaling dimensions of such operators is beyond the scope of our methods.

We can, however, perform a simple analysis of the effect of a small monomer
fugacity at the fixed point.  Consider perturbing the effective Hamiltonian
near the fixed point by a chemical potential $\mu$, which adds a cost for
deviating from the maximum matching condition by introducing monomers. The
effective Hamiltonian with monomer fugacity $\Hc_n'$ at the $n$-th step of RG is given by
\begin{align} \Hc_n' = \Hc_n + N^{n}_m \mu^n,
  \label{eq:monomer_hamiltonian} 
\end{align} 
where $N^n_m$ is the number of
monomers, and $\Hc_n$ is the effective Hamiltonian for dimers studied in Sec.~\ref{sec:effham}. The key point is that by the definition of our coarse-graining
transformation, the number of monomers cannot change under RG; effective dimer
configurations obtained by the transposition procedure described in
Sec.~\ref{sec:rgtrans} preserve the number of monomers, \textit{i.e.},
\begin{equation} 
  \la N^{n+1}_m \ra =\la N^n_m \ra.  \label{eq:monomer_equality}
\end{equation} 
Since the linear dimension of the system decreases by a factor
of $\delta_s^2$ in one step of our RG transformation, near the fixed point we
expect 
\begin{equation} 
  \la \Hc_n' \ra = \delta_s^2 \la \Hc_{n+1}' \ra.  \label{eq:energy_scaling} 
\end{equation}
From Eq.~\eqref{eq:monomer_hamiltonian}, the number of monomers can be
expressed as 
\begin{equation} \la N^n_m \ra = \frac{d\la \Hc_n' \ra}{d \mu^n}.  
  \label{eq:monomer_equation} 
\end{equation}
 Combining Eqs.~\eqref{eq:monomer_equality}, ~\eqref{eq:energy_scaling} and
 ~\eqref{eq:monomer_equation}, it is easy to see that 
 \begin{equation}
   \mu^{n+1} = \delta_s^2 \mu^n.  
 \end{equation}
 We have therefore shown that small perturbations of monomers are
 \emph{relevant} at the fixed point. Since the rescaling associated with the RG
 transformation is exactly $\delta_s^2$, it is clear that the RG eigenvalue is given by
$y_{\mu}=1$.  It might be worth comparing  with the familiar classical dimer models on the square and honeycomb lattices: throughout the critical phase, perturbations in monomer fugacity is relevant, driving the system to a short-ranged fixed point with infinite monomer fugacity; while the monomer fugacity is marginal exactly at the Kosterlitz-Thouless transition to a disordered phase~\cite{Alet_etal}. 

\section{Criticality}
\label{sec:results}

So far, we have established that the dimer model on the \AB~tiling flows to a fixed point under an RG with discrete ``block-spin'' type transformations. We now investigate the nature of the fixed point. A natural question is whether the fixed point Hamiltonian  exhibits critical scaling behaviour. For systems with continuous scale invariance, this manifests itself in the familiar feature that various observables show power-law scaling. Discrete scale invariance modifies this picture: as we  briefly review~\cite{Sorniette}, such power-law scaling forms acquire log-periodic modulations when the relevant fixed points are only invariant under a discrete, rather than continuous, set of scale transformations. 

For a fixed point under an RG transformation $x\rightarrow bx$, one expects, for the singular part of an observable $O(x)$,
\begin{equation}
  O(x)= b^{-\alpha} O(bx).
  \label{eq:o_scaling}
\end{equation}
If the above equation were to hold for all $b$, we have the  familiar power-law solution $O(x) \sim x^{-\alpha}$. However, if Eq.~\eqref{eq:o_scaling}  were only required to hold for a discrete set of scales $b=\lambda^n$ for integer $n$, then substituting $O(x)=x^{-\omega}$ gives us $\omega= \alpha + i 2 \pi k/\log(\lambda)$, for integer $k$. Therefore, the algebraic form $O(x)\sim x^{-\alpha}$ acquires a modulation which is log-periodic in the discrete scale $\lambda$,
\begin{equation}
  O(x)\sim x^{-\alpha}\mathcal{P}_O (\log x /\log \lambda ),
  \label{eq:o_scaling_disc}
\end{equation}
where $\mathcal{P}_O$ is a periodic function with a period of $1$. This log-periodic structure is a distinctive signature of DSI. Therefore, we now attempt to identify such power-law forms with log-periodic modulations for various observables computed from our fixed point Hamiltonian.

\subsection{Dimer correlations}
\label{sec:dimer_correlations}
\begin{figure}[t]
  \includegraphics[width=0.9\columnwidth]{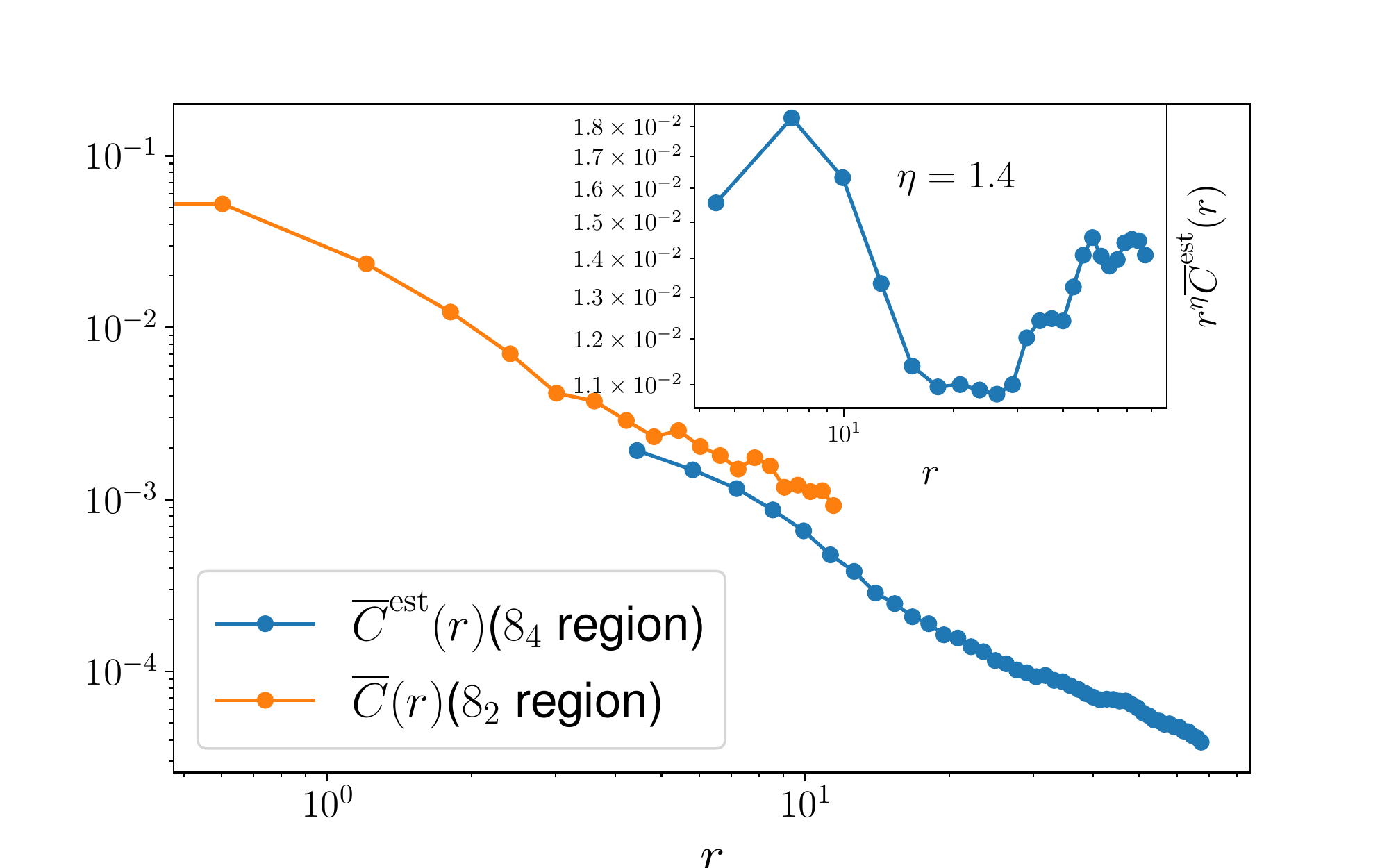}
    \label{fig:corrfunc}
    \caption{ We plot $\ovl{C}(r)$, the average absolute value of dimer correlations for all edge pairs at a separation $r$, calculated from MC simulations of the fixed point Hamiltonian on the $8_2$-region, along with $\ovl{C}^{\mathrm{est}}$, an estimate of $\ovl{C}(r)$ based on the RG fixed point for an $8_4$-region, both on a double-log scale. A power-law behaviour of $\ovl{C}(r)$ modulated by an oscillatory function log-periodic in $\delta_s^2$ is suggested by this.
    The inset shows $\ovl{C}^{\mathrm{est}}$ for an $8_4$-region scaled by $r^{\eta}$, for $\eta=1.4$, suggesting a log-periodic function possibly affected by significant boundary effects.}
\end{figure}
An obvious quantity to investigate is the connected dimer correlations function $C_{ij}$, given by
\begin{equation}
C_{ij} = \langle s(e_i) s(e_j) \rangle - \langle s(e_i)\rangle \langle s(e_j) \rangle
\end{equation}
In earlier work (Ref.~\cite{LloydEA}), it was demonstrated that a subset of dimers have intriguing long-range connected correlations resembling power laws, whereas other dimers have correlations cut-off by lengthscales associated with the effective vertices or $8_n$-regions. The correlation functions do not approximate a smooth function, and  further, do not show any translational or continuous rotational symmetry--- even at large lengthscales. To investigate the nature of decay of these correlations quantitatively, we define $\overline{C}(r)$ which is the average of absolute values of all connected correlations between edge-pairs  whose separations lie within the interval $(r-\epsilon, r+\epsilon)$. The lack of symmetries noted above poses a technical challenge to the numerical calculation of $\overline{C}(r)$: in general, the  calculation of $\overline{C}(r)$ involves the calculation of $N_e^2$ dimer correlations for a system of $N_e$ edges--- a formidable challenge for Monte Carlo calculations in system sizes of interest (for example the $8_4$ region, used in most numerical calculations reported so far, has $N_e=50,000$ edges). 

However, we can make use of the RG transformation near the fixed point to alleviate  this problem.  Consider the free energy $f^n(\{K^n\})=-\log(\sum_{\{s^n_i\}}\exp(-K^n_i s^n_i))$ as a function of the couplings $K^n_i$ at the $n$-th RG iteration. We consider different couplings for all edges $i$ to facilitate calculations of dimer correlations later, even though the fixed point values of $K^n_i$ for many  edges are the same, as explained in Sec.~\ref{sec:effham_results}.
The free energy changes inhomogeneously  under  RG transformations,
\begin{equation}
  f^s(\{K^{n}\})= g(\{K^{n}\}) + f(\{K^{n+1}\}),
\end{equation}
where $g(K)$ is the non-singular contribution to the free energy.
If we are interested only in the singular part of the free energy which controls singular behaviour in all observables including correlation functions, we can ignore the regular part $g(\{K\})$.
The singular  part of the connected dimer correlations is now given by taking second derivatives with respect to the couplings $K_i$ on both sides
\begin{align}
\nonumber  \langle s^n_i s^n_j \rangle^s_c &= \frac{\cd f(\{K^{n+1}\})}{\cd K^n_i \cd K^n_j} \\
\label{eq:corr_func_calc}
  &= \sum_{kl} \frac{\cd K^{n+1}_k}{\cd K^n_i} \frac{\cd K^{n+1}_l}{\cd K^n_j}  \langle s^{n+1}_k s^{n+1}_l\rangle _c
\end{align}
For a system with $N_e$ edges, \eqref{eq:corr_func_calc} uses the RG transformation, encoded in the derivatives
$\frac{\cd K^{n+1}_k}{\cd K^n_i}$, to calculate $N_e^2$ correlation functions on the LHS in terms of $\frac{N_e}{\delta_s^4}^2$ 
correlation functions in the RHS. The derivatives can be calculated using MCRG, by solving a chain-rule equation [see also Eq.~\eqref{eq:mcrg_chainrule} in Appendix~\ref{app:dimer_irrelevance}],
\begin{equation}
  \langle s^{n+1}_i s^n_j \rangle_c = \sum_k \frac{\cd K^{n+1}_k}{\cd K^{n}_j} \langle s^{n+1}_i s^{n+1}_k\rangle_c.
\end{equation}

We use this technique to make estimates of $\overline{C}(r)$, which we call $\overline{C}^{\text{est}}(r)$, for the $8_4$ region, which affords access to sizes $r\sim100$. In Fig.~\ref{fig:corrfunc}, we display $\overline{C}^{\text{est}}(r)$  along with $\overline{C}(r)$ calculated directly from MC samples for the modestly-sized $8_2$ region where such calculations are still feasible. The data appears consistent with a power-law modulated by log-periodic function. In the inset, we show $r^{\eta} \overline{C}^{\text{est}}(r)$ with $\eta\sim 1.4$, which shows an approximately periodic function on the logarithmic scale. While this is suggestive of an underlying DSI critical point with $\overline{C}(r) \sim r^\eta$, the second period is already dominated by boundary effects. Since larger sizes remain inaccessible to the calculation of $\overline{C}^{\text{est}}(r)$, systematic use of finite-size scaling theory to conclusively establish  criticality and determine the exponent $\eta$ is not currently within the reach of our numerics. We instead consider an alternative approach to probing the criticality and DSI at the fixed point, using the structure of the structure of overlap loops of two decoupled dimer models.

\subsection{Random geometry of overlap loops in the double dimer model}
\label{sec:loopmodel}
\begin{figure}[t]
    \includegraphics[width=\columnwidth]{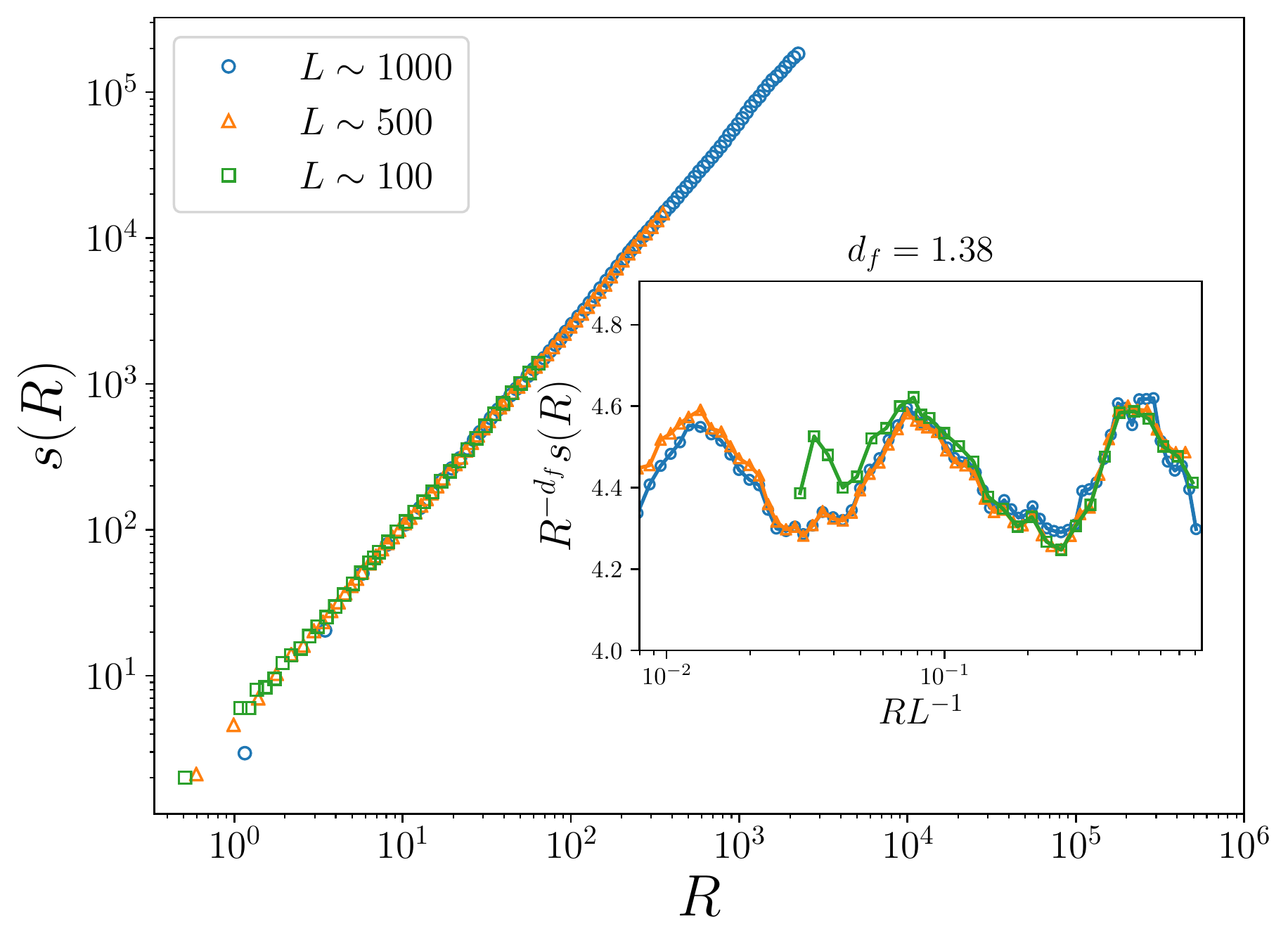}
    \caption{The average size $s(R)$ of a loop plotted against its radius $R$ reveals a power-law with (weak) periodic corrections on the log-log scale.  Plotting $s(R)R^{-d_f}$ against $R/L$, the data for different sizes collapse to a finite-size scaling form (Eq.~\eqref{eq:fracdim_scaling2}) for a fractal dimension $d_f=1.38$, revealing the weak log-periodic modulations to power-law behaviour. }
    \label{fig:frac_dim}
\end{figure}

\begin{figure}[t]
    \includegraphics[width=\columnwidth]{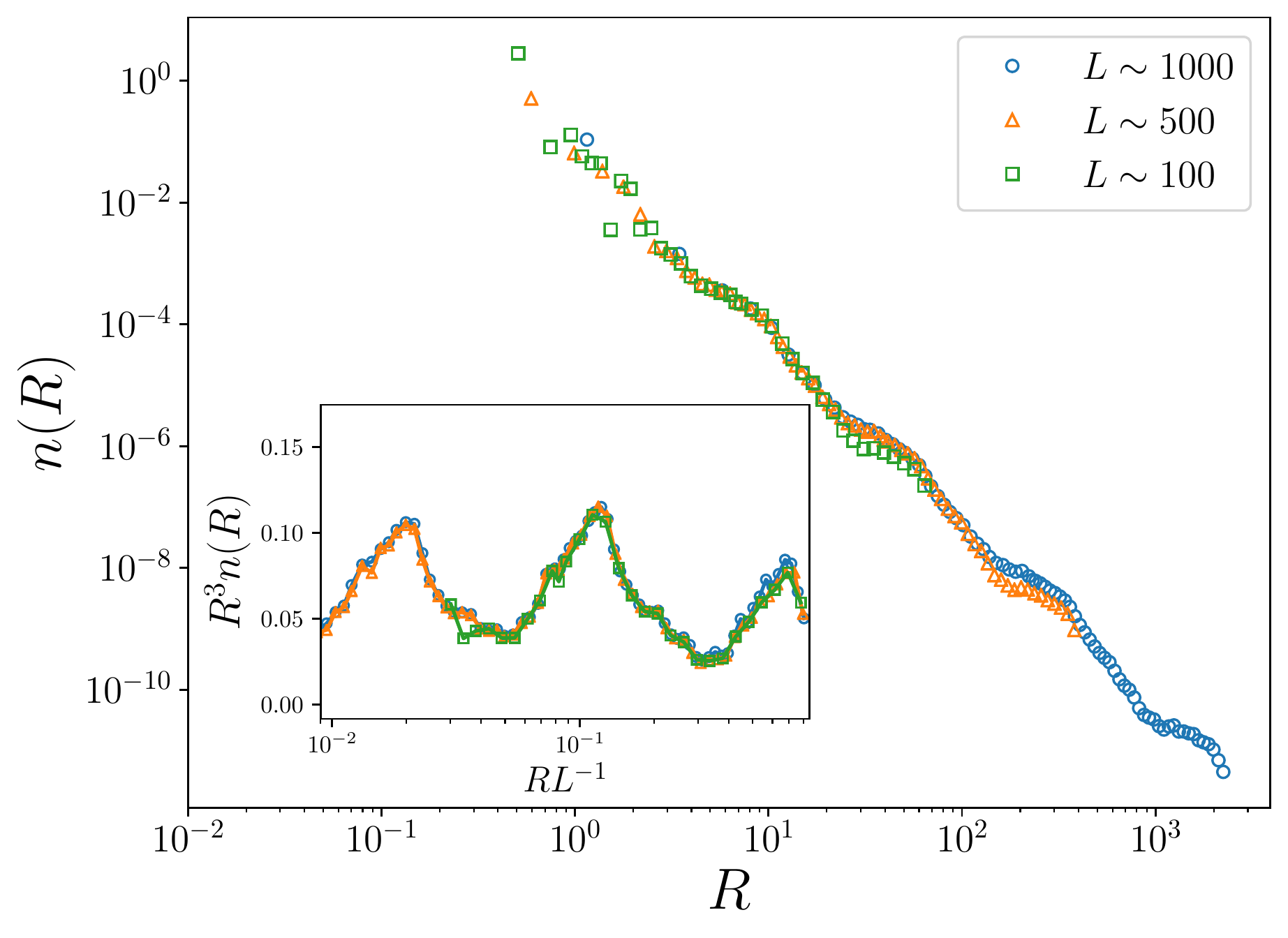}
    \caption{The average number density $n(R)$ of a loop plotted against its radius $R$ reveals a power-law with periodic corrections on the log-log scale.  Plotting $n(R)R^{3}$ against $R/L$, the data for different sizes collapse to a finite-size scaling form (Eq.~\eqref{eq:nr_scaling2}), revealing the log-periodic modulations to power-law behaviour. A collapse with an exponent of $3$ (and no other free parameters) confirms the expectations of the scaling theory presented in the text.}
    \label{fig:nr}
\end{figure}
\begin{figure}[t]
    \includegraphics[width=\columnwidth]{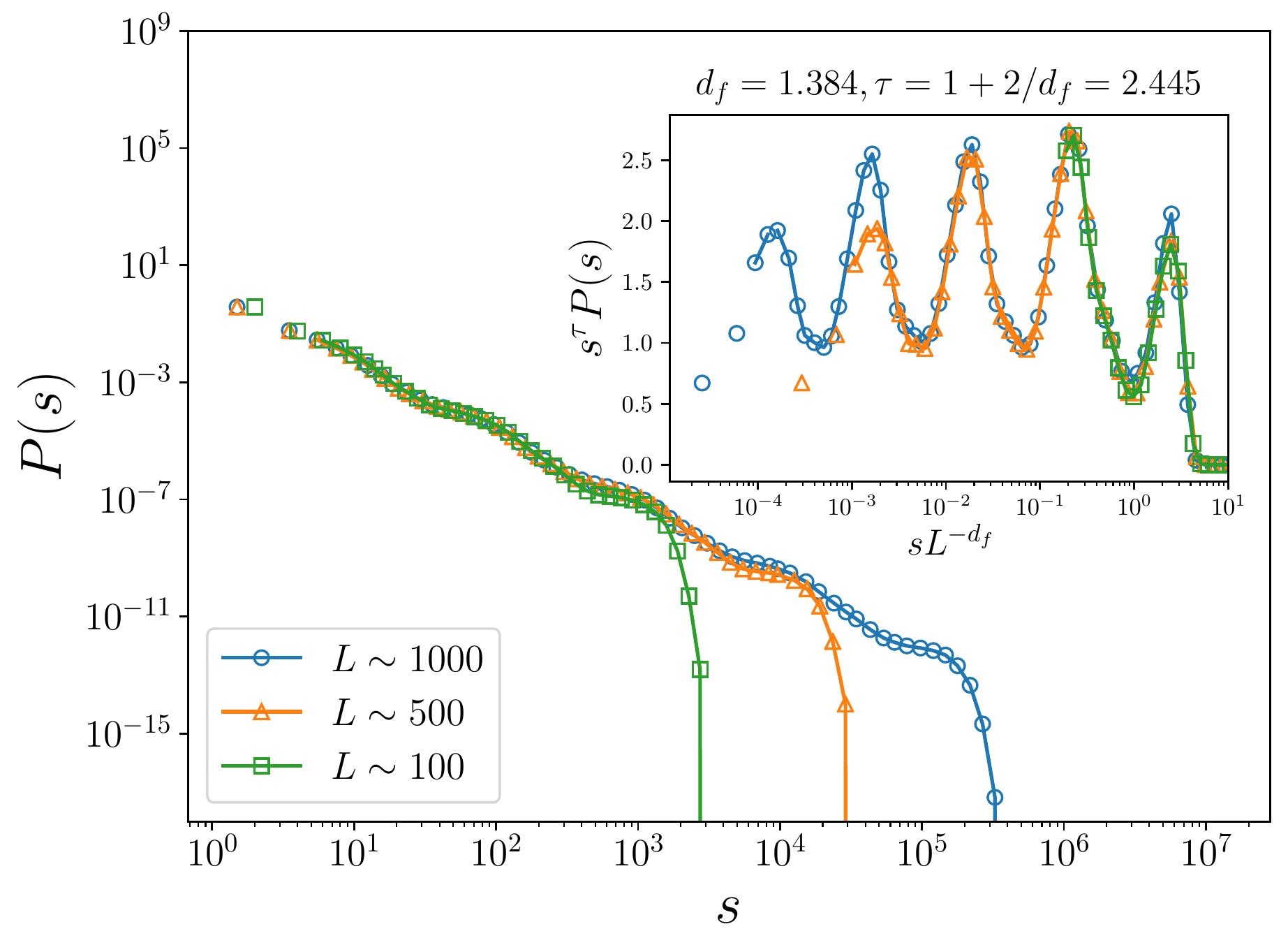}
    \caption{$P(s)$, the density of loops of size $s$  plotted against $s$ also reveals a power-law with periodic corrections on the log-log scale. Plotting $P(s)s^{\tau}$ against $s^{-d_f}/L$, the data for different sizes collapse to a finite-size scaling form (Eq.~\eqref{eq:ps_scaling2}) for $\tau=2.55$ and $d_f=1.37$, from Fig.~\ref{fig:frac_dim},  revealing the log-periodic modulations to power-law behaviour. The exponent $\tau$ and $d_f$ used in the scaling collapses are consistent with the scaling relation $d_f (\tau-1)=2$, expected to hold from the scaling theory presented in the main text.}
    \label{fig:ps_scaling}
\end{figure}
To quantitatively probe criticality at the DSI fixed point, we  consider the seemingly-unrelated problem of a {\it double dimer model}. This is constructed by
superposing two independent dimer covers, which defines a configuration of tightly-packed self-avoiding loops on the \AB~graph, with the additional possibility that links can host
loops of length 2, corresponding to the overlapping of dimers at the same edge from both covers. Thus, the double 
dimer model can be mapped to a loop gas. There is a long history of the investigation of such loop gases, whose properties are intimately connected with those of the underlying dimer model. For instance, for bipartite dimer
models with a height description (e.g. on periodic graphs), the long-wavelength properties of  loops in the double dimer model correspond to equal-height contours of the fluctuating
height field~\cite{KondevHenley_fourcoloring,DesaiEA}, and have been investigated in detail with a scaling theory~\cite{KondevHenley, KondevHenleySalinas}. In such cases, the ensemble of overlap loops is also rigorously known to be conformally invariant~\cite{Kenyon_doubledimer}. 

Our primary motivation for studying the double dimer model and the associated loop gas is the relative numerical accessibility of loop observables in Monte Carlo simulations.  Incontrovertible evidence of critical scaling of the loop observables, defined in the configuration space of two decoupled dimer models, provides a very strong suggestion of criticality in the single dimer model.  
To test the critical scaling of this loop gas,  
we use the scaling theory of the critical loop ensembles, largely following Kondev and Henley~\cite{KondevHenley}, but additionally modifying their power-law scaling {\it ansatzes} with log-periodic modulations appropriate for DSI.  We present numerical evidence of DSI in the loop gas by showing that Monte Carlo calculations of observables of the loop gas are consistent with such a critical scaling theory of loops with log periodic modulations.

Following the motivations outlined at the beginning of Sec.~\ref{sec:mcrg}, we choose effective vertices, or $8_n$ regions, as our samples, specifically
the $8_4$, $8_6$ and $8_8$ regions. They have $15473$, $517825$ and $17520593$ vertices respectively. In terms of units of edge separation, the linear dimensions are  $L\sim 100$, $L\sim 500$, and $L\sim 3000$ respectively. In these simulations, we use the fixed-point values of dimer weights obtained from the RG calculations described in Sec.~\ref{sec:rgtrans} and Sec.~\ref{sec:effham}

For each loop, we define a loop size $s$ (not to be confused with dimer occupancy variables defined earlier) and a loop radius $R$. The loop size is the number of edges in a loop, while $R$ is operationally defined as the radius of the smallest disk which completely covers the loop. At a critical point with scale symmetry at discrete scales given by $\delta_s^{2n}$, from Eq.~\eqref{eq:o_scaling_disc} we expect 
\begin{equation}
  \label{eq:fracdim_scaling1}
  s(R) = R^{d_f} \mathcal{P}_{d_f}(\log R/\log \delta_s^2 ),
\end{equation}
where $\mathcal{P}_{d_f}(\log R/\log \delta_s^2)$ is a log-periodic function.  Our finite-size scaling hypothesis for $s$  is 
\begin{equation}
  \label{eq:fracdim_scaling2}
  s(R,L) = R^{d_f} \mathcal{F}_{d_f}(RL^{-1}),
\end{equation}
where we expect the scaling  function $\mathcal{F}_{d_f}(RL^{-1})$ to be equal to  $\mathcal{P}_{d_f}(\log R/\log \delta_s^2)$ for $RL^{-1} \ll 1$, and encode the  effects of finite sample size and boundary termination for $RL^{-1} \simeq 1$. 
We expect systems of different linear dimensions $L$ to collapse to the same scaling function $\mathcal{F}_{d_f}$ only if these systems are related by discrete scale transformations. Our samples, the $8_4$, $8_6$, and $8_8$ regions, satisfy this condition. 
We display a histogram of loop sizes $s$ plotted against their radius $R$ in Fig.~\ref{fig:frac_dim}. We see that the histograms approximate a power-law with periodic modulations of very small magnitude, establishing the fractal nature of our loops. We can perform a finite-size scaling collapse (Eq.~\eqref{eq:fracdim_scaling2}) on plotting $sR^{-d_f}$ against $RL^{-1}$ with the exponent $d_f=1.384(1)$. The error bars reflect the range of $d_f$ over which a good collapse can be obtained. The scaling function clearly demonstrates the log-periodic modulations which are the hallmark of DSI.

Next, we consider the quantity $n(R)$,  the density of loops with radius in $(R,R+dR)$. If we consider a patch of area $A$, the number of loops in $(R,R+dR)$ in the patch is 
then $n(R) A \,dR$. Under a coarse-graining transformation $R\rightarrow b^{-1}R$ with $b=\delta_s^2$, the same loops lie in a patch of area $b^{-2}A$ with radii in $(b^{-1}R,b^{-1} (R+dR))$ . If we have DSI, then the same function $n(R)$ describes the rescaled loop ensemble, and so we have  $n(R) A\,dR$ = $n(Rb^{-1})A\,dR\, b^{-3}$. Following Eq.~\eqref{eq:o_scaling_disc}, we then have the scaling form \footnote{Note that general loop ensembles often have $n(R)\sim R^{-3 + \alpha}$ with nonzero $\alpha$, \textit{e.g.}, contour loops of fluctuating rough surfaces.}
\begin{equation}
  \label{eq:nr_scaling1}
  n(R) = R^{-3} \mathcal{P}_{n}(\log R/\log \delta_s^2 ),
\end{equation}
and the corresponding finite-size scaling form
\begin{equation}
  \label{eq:nr_scaling2}
  n(R,L) = R^{-3} \mathcal{F}_{n}(RL^{-1}).
\end{equation}
As usual, we expect $\mathcal{F}_{n}$ to be equal to the periodic function $\mathcal{P}_{n}$ when  $RL^{-1} \ll 1$ and incorporate finite size effects when $RL^{-1} \simeq 1$. We show, in Fig.~\ref{fig:nr}, that plotting $n(R,L) R^{3}$ against $RL^{-1}$  gives us an excellent scaling collapse, without any free parameters, into a function periodic in $\log \delta_s^2$--- thereby providing strong evidence of critical scaling and DSI.
 
Finally, we look at the quantity $P(s)$, the area density of loops of size between $s$ and $s+ds$. We use a similar argument  about the total number of loops in a patch of area $A$ remaining invariant under the rescaling $R\rightarrow Rb^{-1}$ with $b=\delta_s^2$, which implies$P(s)A\,ds=P(s/b^{d_f})A \,ds\, b^{-2-d_f}$, giving  the scaling forms
\begin{align}
  \label{eq:ps_scaling1}
  P(s)\sim s^{-\tau} \mathcal{P}_P (\log s/ \log \delta_s^{2 d_f}), \text{ and} \\
  \label{eq:ps_scaling2}
  P(s,L) = s^{-\tau} \mathcal{F}_{P}(sL^{-d_f}).
\end{align}
The finite-size scaling function $\mathcal{F}_P$ is equal to the log-periodic function $\mathcal{P}_P$ for $sL^{-d_f} \ll 1$, and the exponent $\tau$ satisfies the scaling relation
\begin{equation}
  \label{eq:scaling_relation}
  d_f(\tau-1)=2.
\end{equation}
We first plot $P(s)$ against $s$ in Fig.~\ref{fig:ps_scaling} to demonstrate a power-law with log-periodic modulations, in the inset we show that plotting $P(s) s^{\tau}$  against the scaling variable $sL^{-d_f}$, with $d_f=1.384$ obtained above and $\tau=1+2/d_f$ from Eq.~\eqref{eq:scaling_relation} obtains a good scaling collapse. We note that all scaling collapses (Figs.~\ref{fig:frac_dim}, ~\ref{fig:nr} and ~\ref{fig:ps_scaling}) were obtained with a single free parameter, the fractal dimension $d_f$ of the critical loops.

Finally, one might consider the loop correlation function $G(r)$ which is the average probability that two sites separated by a distance $r$ is a part of the same loop. For usual critical loop ensembles~\cite{KondevHenley}, one expects $G(r)\sim r^{-2x}$, with the exponent $x$ satisfying the scaling relation $d_f=2-x$. We expect similar forms to hold for our case, albeit with the usual log-periodic modulations. Calculating loop-correlation functions are numerically challenging, and we do not attempt to do so here. However our scaling analyses already predict $G(r)\sim r^{-1.23}$ from the scaling relation relating $x$ to the fractal dimension of the loops. Therefore, by studying the ensemble of overlap loops, we have established that two decoupled copies of classical dimer models are  critical in an unusual way, exhibiting discrete instead of continuous scale invariance. This strongly suggests that the underlying classical dimer model also has similar critical properties.

\section{Discussion}
\label{sec:discussion}
We have constructed a numerical real-space RG transformation for the classical dimer model on the AB tiling. Implementing these transformations by large-scale Monte Carlo simulations, we show that our model flows to a fixed point under discrete ``block-spin'' type RG transformations. We have also introduced a Monte-Carlo-based technique to calculate the effective Hamiltonians, using  it to track RG flows and explicitly construct the fixed point Hamiltonian in terms of dimer weights on different types of edges. The ability to write down and simulate the fixed point Hamiltonian 
directly has allowed us to make further progress: most importantly, we show that dimer correlation functions on accessible system sizes are consistent with the expectation of power-laws modulated by log-periodic corrections --- the hallmark of a critical point with DSI. Finally, we studied the loop ensemble defined by the overlap graphs of two-decoupled dimer models; loop observables in this auxiliary ensemble are numerically more accessible, and this allows us to show conclusive evidence of critical scaling with DSI, via the explicit computation of power laws with log-periodic modulations which are in perfect agreement with a DSI-adapted scaling theory of the critical loop ensemble.

Very recent work has discovered a DSI critical point (with similar lack of emergent translational and rotational symmetries) in a  quasiperiodic percolation problem~\cite{SommersEA}, 
in which the percolating bonds are chosen according to a quasiperiodic pattern.
DSI has also been observed before in critical phenomena. Most such examples, however, involve quantum spin chains, where older results~\cite{LuckXY,HermissionEA,Hermission, Hida, Vieira,SatijaEA} focussed on  quasiperiodic couplings generated by certain binary substitution rules have been recently supplemented by results on more general quasiperiodic modulations~\cite{CrowleyEA1,CrowleyEA2,AgrawalEA1,AgrawalEA2}.  DSI has also been reported in classical statistical mechanics~\cite{Derrida1984,Andrade}, albeit on heirarchical graphs. These graphs are closely 
connected to real-space renormalisation transformations, often defined such that RG transformations of the  Migdal-Kadanoff type are exact on them by construction~\cite{Kaufman_Griffiths1}. Hierarchical graphs are unusual in terms of their geometry, and not straightforwardly generalisable to other systems. While quasiperiodic systems such as those considered in this paper have a built-in scale symmetry (and in this sense resemble hierarchical graph) they are typically not amenable to exact real space block-spin RG~\cite{penrose_rsrg,ab_rsrg}. Indeed,  DSI is only manifest Monte Carlo renormalisation procedure. To our knowledge, other statistical mechanical models with uniform couplings on quasiperiodic graphs host conventional critical phenomena with continuous scale invariance~\cite{IsingPenrose,IsingAB,PenroseXY, PottsAB, PottsAB_2}. Further,
 the log-periodic modulations acquired by power-laws at  critical points of hierarchical graphs are typically  very weak~\cite{Derrida1984} --- typically a factor of $\sim 10^{-5}$ weaker than the background ---  whereas in the present example  DSI is clearly manifest at the critical point.

 Another, possibly related feature of our model is that the theory at the fixed point is described
 by hard-core dimers. This is in contrast to critical points of constrained dimer and loop models which are described by continuum fields. This has  been confirmed by parallel work~\cite{GoekmenEA} using machine-learning-based coarse-graining schemes~\cite{Koch-JanuszEA,GoekmenEA}. These schemes have ``discovered'' that the effective degrees of freedom are hard-core dimers~\cite{GoekmenEA2} based purely on information theoretic considerations--- \textit{i.e.}, without any prior knowledge about the structure of perfect matchings. 

 Some future directions are naturally suggested by our work. The first concerns loop models. The partition  function of the double dimer model  considered here, expressed in terms of the overlap loops, is given by $Z=\sum_{\mathrm{C}} 1^{N_d}2^{N_L}$, where $N_d$ is the number of doubled edges  with two dimers (one from each copy of the dimer model), and $N_L$ is the total number of other non-trivial loops. These are closely
 connected to the $O(n)$ loop models~\cite{Blote_Nienhuis,BatchelorEA_1996}, that describe a configuration of
 closely packed loops with $Z=\sum_{\mathrm{C}}n^{N_L} $; the configuration space typically excludes doubled edges. The AB~graph has been recently shown to be perfectly packed by such loops~\cite{SingLlyodFlicker}. The $O(2)$ loop model is expected to have similar critical behaviour to the double dimer model  considered here, since the weights for all non-trivial loops are the same in both models. Since we have shown that the double dimer model is critical, $O(n)$ loop models with $n \leq 2$, \textit{i.e.,} ones with larger relative weight for longer loops, are  expected to be either critical or in the long-loop phase. It is clear that for large $n$ where long loops are suppressed, loop models must be in a short-loop phase.  The question then is to understand the critical behaviour of $O(n)$ models as a function of $n$ to investigate a transition from a critical or long-loop phase to a short-loop phase. (For $O(n)$ loop models on square and honeycomb lattices, the model is critical for $n \in [0,2]$, and short-ranged for $n > 2$~\cite{Blote_Nienhuis}.) For periodic bipartite lattices, other related models such as the non-crossing~\cite{RaghavanHenley} and bilayer~\cite{WilkinsPowell,DesaiEA} dimer models and  colouring problems~\cite{KondevHenley_fourcoloring} also show a wealth of interesting critical behaviour that can be understood in terms of height variables; the fate of such models on AB graphs remains an important open question.

 A second direction concerns the properties of nearest neighbour Resonating Valence Bond (nnRVB) wavefunctions  on these graphs. nnRVB wavefunctions often exhibit a rich host of spin-liquid behaviours and algebraic correlations~\cite{TangSandvikHenley,AlbuquerqueAlet,PatilEA,AlbuquerqueEA}. RVB wavefunctions for $SU(g)$-spins can be characterised by an overcomplete basis of dimer configurations, with each dimer representing an $SU(g)$-singlet~\cite{Sutherland1988,BeachSandvik,BeachEA}. Observables of such RVB wave-functions are given by estimators of overlap loops of dimer configurations, with distributions given by the partition function $Z(g)=\sum_{\mathrm{C}} g^{N_d} (2g)^{N_L}$. In particular, spin-correlation functions $\langle\vec{S}_j\vec{S}_j \rangle$ are related to the loop-correlation functions evaluated in $Z(g)$. On a square lattice, $Z(g)$ is short-ranged for the $g=2$ case relevant to $SU(2)$-spins, resulting in short-range spin correlations. However the valence-bond correlations $ \langle P_{ij}P_{kl} \rangle$, with the projector $P_{ab}=\langle \vec{S}_a \cdot \vec{S}_b -1/4\rangle$, are still critical~\cite{TangSandvikHenley,AlbuquerqueAlet}. This can be understood from the observation that such valence bond correlations, in the large-$g$ limit, are equal to the classical dimer correlations~\cite{RamolaEA}, which are known to be critical for the square and honeycomb lattices. The fateof such nnRVB wavefunctions for $SU(2)$-spins  is still an open question for the AB~graph: even if spin-correlations are short-ranged like the square and honeycomb lattice cases, nnRVB wavefunctions can still host critical valence-bond correlations, as we have shown in Sec.~\ref{sec:dimer_correlations} that the classical dimer-correlations appear critical. In this light, investigating properties of nnRVB wave functions is likely to be a rewarding enterprise.

 A final direction  is to try to the drive the dimer model out of criticality. A first step in this direction might incorporate  aligning interactions~\cite{Alet_etal}, and study how their presence affects the RG flow. 
 
 We close by noting that the AB~graph can be deformed into closely-related cousins which are likely to be numerically more accessible while sharing similar properties. $D_8$-symmetric samples of the AB~graph, such as those considered in most numerical investigations of this paper, have been argued to control the structure of maximum matchings in all samples in thermodynamic limit.  These samples are built of of eight equivalent sectors. Decreasing the number of sectors leads to smaller graphs (though no longer tilings), while preserving the scale-invariant structure of perfect matchings. Whether classical dimer models on these deformed graphs leads to DSI fixed points remains an open question. We expect such deformations to be useful in numerically attacking questions on quantum-mechanical dimer models, as well as in performing quantitative calculations of critical behaviour of dimer and monomer correlations of the classical dimer model which remain inaccessible even to our large-scale numerics on the $AB$-graph.

\acknowledgements
We  are grateful  Tyler Helmuth and Kedar Damle for suggesting that we investigate the double dimer model, and Abhishodh Prakash and Michele Fava for crucial discussions throughout the period of the project. We thank  Jerome Lloyd, Steven Simon, and Felix Flicker for a previous collaboration studying the AB tiling  (Ref.~\cite{LloydEA}). SB thanks  Doruk G\"{o}kmen, Sebastian Huber, Maciej Koch-Janusz, Zohar Ringel, and Felix Flicker for collaboration on recent related work (Ref.~\cite{GoekmenEA2}).  We acknowledge support from the European Research Council under the European Union Horizon 2020 Research and Innovation Programme via Grant Agreement No. 804213-TMCS.

\appendix
\section{Fixed point behaviour of dimer correlations}
\label{app:dimercorrs}
Earlier, in Tab.~\ref{tab:dim-den}, we showed how effective dimer densities of coarse-grained dimer models on an $8_1$-region flows to a fixed point in the limit of  effective densities being obtained by coarse-graining large microscopic dimer models. Specifically, we consider $8_{1+2n}$-regions and coarse-grain them to $8_{1}$ regions,  measuring the effective dimer densities and checking that they flow to a fixed point with $n$.
For a fixed point, we need not only the dimer densities, but the whole probability measure of dimer distributions to go to a fixed point with increasing $n$. Here, in Tab.~\ref{tab:dim-corr}, we show how connected correlations of dimers also flow to a fixed point under RG transformations. While we display the data
for a particular choice of source edge, ($e_j=3$ in the labelling convention of Fig.~\ref{fig:8_1_region}), the convergence to the fixed point has no dependence on this choice.

\begin{table}[h]
  \centering
  \begin{tabular}{ |p{0.08\columnwidth}| p{0.17\columnwidth}| p{0.17\columnwidth} | p{0.17\columnwidth}| p{0.17\columnwidth}|}
  \hline
 Edge&\multicolumn{4}{|c|}{Effective dimer correlations after $n$ RG steps} \\
\hline
    & n=0 & n=1 & n=2 & n=3 \\
\hline
\hline
0 & 0.2021(1) & 0.2058 & 0.2066 & 0.2070(1) \\
\hline
1 & -0.1351(1) & -0.1424 & -0.1434 & -0.1437(1) \\
\hline
2 & 0.1347(1) & 0.1422 & 0.1432 & 0.1434(1) \\
\hline
3 & -0.0792(1) & -0.0839 & -0.0851 & -0.0853(1) \\
\hline
4 & -0.0670(1) & -0.0616 & -0.0617 & -0.0618(1) \\
\hline
5 & -0.0555(1) & -0.0564 & -0.0566 & -0.0567(1) \\
\hline
6 & -0.0582(1) & -0.0562 & -0.0561 & -0.0562(1) \\
\hline
7 & -0.0582(1) & -0.0563 & -0.0560 & -0.0561(1) \\
\hline
8 & 0.0490(2) & 0.0474 & 0.0475 & 0.0476(1) \\
\hline
9 & 0.0405(1) & 0.0434 & 0.0435 & 0.0435(1) \\
\hline
10 & 0.0177(1) & 0.0149 & 0.0148 & 0.0147(1) \\
\hline
11 & 0.0147(1) & 0.0137 & 0.0136 & 0.0136(1) \\
\hline
12 & 0.0151(1) & 0.0133 & 0.0131 & 0.0131(1) \\
\hline
13 & 0.0153(1) & 0.0133 & 0.0131 & 0.0129(1) \\
\hline
14 & -0.0125(2) & -0.0100 & -0.0099 & -0.0099(1) \\
\hline
15 & 0.0124(2) & 0.0098 & 0.0097 & 0.0098(1) \\
\hline
\end{tabular}
\label{tab:dim-corr}
\caption{Starting with the graph $\Gc$ being an $8_1$ region of Fig.~\ref{fig:8_1_region}, we consider a series of inflated graphs $\Gc_n=\sigma^{2n}$ for $n=0,1,2,3$. We use our RG transformation $n$ times for the graph $\Gc_n$ to coarse grain it back to the $8_1$-region $\Gc_0$ and compute effective dimer correlations there. In labelling for symmetry-inequivalent edges introduced in Fig.~\ref{fig:8_1_region} to report the dimer correlations 
$\langle s(e_i) s(e_j)\rangle -\langle s(e_i) \rangle \langle s(e_j)\rangle$ for $e_i=3$, and sort $e_j$ in decreasing order of magnitude of correlations.
We see that they reach a fixed point with increasing $n$, implying that the coarse-grained dimer problem on the $8_1$-region has approached a fixed point. This complements Tab.~\ref{tab:dim-den} in the main text, which tracks dimer densities as they flow to a fixed point.  }
\end{table}

\section{Alogrithm to calculate effective dimer weights}
\label{app:algorithm}
For concreteness, let us consider the effective configuration $\{s^1_i\}$ on a
graph $\Gc_1$, obtained by coarse-graining (Eq.~\eqref{eq:rg_transf1})
a microscopic configuration $\{s^{ 0}_i\}$  on the graph $\Gc_0$ We will now describe a Monte Carlo procedure to
estimate the weights of effective dimers $W_1(e_k)$.  We start with the graph $\Gc_0$, and add to it all the edges which
belong to the coarse-grained graph $\Gc_1$, with dimer weights on
these new edges set to 1. The dimers on these new edges will be called
``long-range dimers". On the configuration space of this auxiliary
dimer model we impose the additional constraint that there can be at most one
long-range dimer. If we now coarse-grain a configuration with a long-range
dimer on the edge $e_k$ by transposing with a dimer configuration on the
corresponding \ABB~graph as before, the transposition graph will have open
alternating paths connecting all the 8-vertices, except the pair  corresponding
to the edge $e_k$ which already hosts the long-range dimer. Since the
long-range dimer has a weight of $1$, such configurations approximately sample
the partition function described by the denominator of Eq.~\eqref{eq:wt_calc}.

Before we convert this into an
operational algorithm for calculating dimer weights, we must address another
subtlety. In the context of the RG transformations, the effective dimers
correspond to open paths connecting 8-vertices obtained from transposition
graphs of the double ensemble of matchings in \AB~and the corresponding
~\ABB~graphs. In this context, the partition function of the auxiliary dimer model with a single long-range
dimer fixed on $e_k$ is approximately equal to the partition function of the effective dimer model obtained by
fixing not only the effective dimer on the edge $e_k$, but also the specific open
path in the transposition graph which resulted in the effective dimer; the
entropy of the choice of open path is captured by the weight $W_1(e_k)$. The
long-range dimer, then, is a proxy for a single open path corresponding to an
effective dimer on the edge $e_k$. Crucially, the open paths obtained from
transposition graphs obtained from the double ensemble do not intersect each
other (by construction); they have an exclusion constraint. Our proxy
long-range dimer on the edge $e_k$ of our auxiliary dimer problem, however, is
completely permeable to open paths between other $8$-vertices corresponding the
other effective dimers. We rectify this problem approximately: we maintain a
precomputed list of ``shortest open paths" for each edge $e_k$, and calculate
the partition function with the long range dimer fixed at edge $e_k$ and the
additional constraint that the open paths corresponding to other effective
dimers do not intersect the shortest path corresponding to the edge $e_k$. 

We distill the main ideas from the previous paragraphs into an algorithm to calculate weights of effective dimers 
$W_{n+1}(e_{k})$ on the graph $\Gc_{n+1}$, given the dimer weights $W_{n}(e_k)$ on the 
graph $\Gc_n$:
\begin{enumerate}
  \item Use weights $W_n$ to Monte Carlo-sample dimer configurations $\{s^n_i\}$ on the graph $\Gc_n$. Independently sampling dimer configurations on the corresponding \ABB~graph $\Gc_{n/n+1}$, use the double ensemble to construct coarse-grained dimer configurations $s^{n+1}_{i}$ on the graph $\sigma^{-2(n+1)}(\Gc)$, as described in Sec. ~\ref{sec:rgtrans}. Calculate the dimer densities $\langle s^{n+1}_k \rangle$ on all edges.
  \item For each edge $e_k$ in $\Gc_{n+1}$, compute $\Pcs(e_k)$, the shortest path between
    the 8-vertices in $\Gc_n$ which correspond to the edge $e_k$.
  \item Starting from the graph $\Gc_n$,  add edges present in $\Gc_{n+1}$. Construct an auxiliary dimer model on this graph as follows: All dimers on the edges in $\Gc_n$ have  weights given by $W_n$ as before, while the ``long-range" dimers between 8-vertices have weights of 1. Additionally,  this new dimer model has the constraint that there is at most 1 long-range dimer present in an allowed configuration.
  \item Monte Carlo-sample the auxiliary dimer model described above. Calculate the fraction of configurations without long-range
    dimers  $\bar{\rho}$. If a configuration has a long-range dimer on the edge $e_k$, we sample a dimer configuration on the \ABB~graph corresponding to $\Gc_{n/n+1}$. Now we coarse-grain, by transposing these two dimer configurations, leading to open paths linking up all 8-vertices in pairs, except the pair corresponding to the edge $e_k$. If none of these open paths intersect the precomputed path $\Pcs(e_k)$, we add to a buffer. The ratio of the number of configurations in this buffer to the total number of configurations gives us  $\langle \rho_k \rangle$.
  \item The estimator for the weight $W_{n+1}(k)$ is given by 
    \begin{equation}
      W_{n+1} (k) = \frac{\langle s^{n+1}_k \rangle}{  \langle \rho_k \rangle / \bar{\rho} }
  \end{equation}
\end{enumerate}
\section{Details on effective Hamiltonian approximations}
\label{app:effham_details}
\begin{figure}[t]
    \includegraphics[width=\columnwidth]{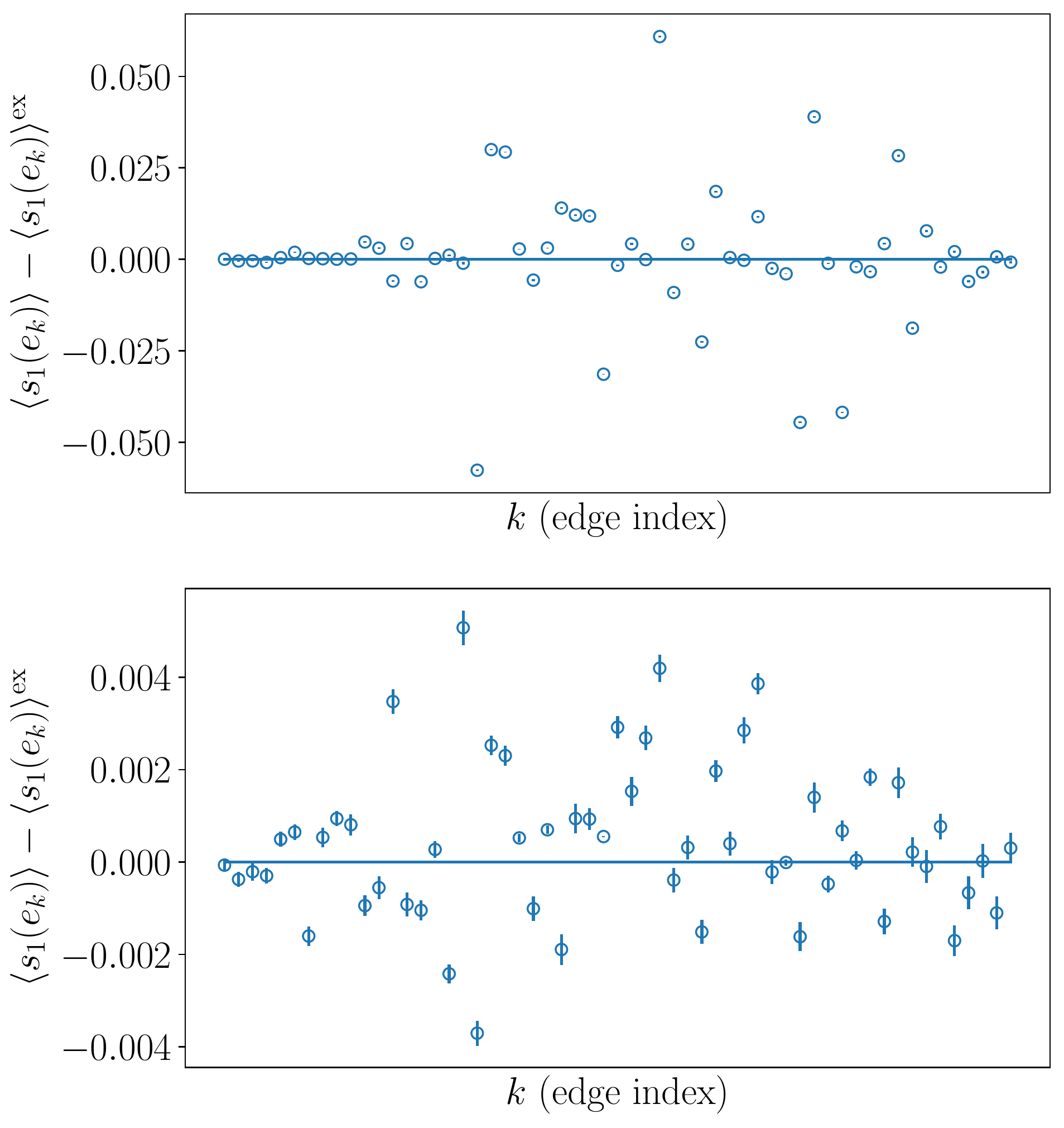}
    \caption{We display and compare the error accumulated by dimer-densities calculated using the approxmations $\ovl{\Hc}$ and $\oo{\Hc}$ to the effective Hamiltonian obtained by coarse-graining the microscopic dimer problem on an $8_4$-region to an effective problem on an $8_2$-region.  Top (Bottom): The difference of effective dimer-densities $\ovl{\langle s_1(e_k) \rangle}$ ($\oo{\langle s_1(e_k) \rangle}$), calculated with the MC simulations of the approximate effective Hamiltonian $\ovl{\Hc}$ ($\oo{\Hc}$)  on an $8_2$-region, to the exact  effective dimer-densities $\langle s_1(e_k)\rangle_{\mathrm{ex}}$, calculated directly from MCRG coarse-graining of an $8_4$-region.}
    \label{fig:compare_effham1}
\end{figure}

In the main text, we hypothesized that the effective Hamiltonian involves effective dimers getting weights $W_n(e_k)$ on all edges $e_k$. An effective Hamiltonian thus obtained, $\Hc_n$ has as one independent parameters for each symmetry-inequivalent edge of the graph.

We proposed two approximations to the effective Hamiltonian that reduce the number of free parameters to an $O(1)$ number. The first, $\overline{\Hc_n}$, is based on the observation that the weight $W_n(e_k)$ depends strongly on the type of the edge, $T(e_k)$, and approximates $W_n(e_k)$ by $\overline{W_n}(t)$, the average $W_n(e_k)$ over all edges $e_k$ with type $T(e_k)=t$.
The second, $\oo{\Hc}_n$, is based on incorporating the effect that the alternating paths which correspond to effective dimers in the RG transformation cannot intersect, and consequently might result in an interaction between effective dimers on parallel edges. To account for this, $\oo{\Hc}$ allows the weight to depend not only on the edge type $t$, but also $t^{||}_1$ and $t^{||}_2$, the types of the two edges parallel to $e_k$. It replaces $W_n(e_k)$ by $\oo{W_n}(t,t^{||}_1,t^{||}_2)$, the average weight over all edges such that their edge-type is $t$ and  they have parallel edges with edge-types $t^{||}_1$ and $t^{||}_2$.

In this Appendix, our goal is to check that these are reasonable approximations; \textit{i.e.}, to
verify that the partition function of Eq.~\eqref{eq:partition_func}  with the
effective Hamiltonian calculated according to these approximations correctly reproduces
the effective dimer distributions obtained directly via the RG
transformation. 
To check these effective Hamiltonians, we first calculate the effective dimer
densities  \emph{exactly} on an $8_2$-graph by coarse-graining
dimer distributions on an $8_4$ graph using MCRG; we denote this by $\langle
s_1(e_k) \rangle^{\mathrm{ex}}$, where the subscript $1$ denotes the fact that the
effective dimer densities are obtained after 1 RG step. We compare this against
the dimer densities calculated by MC sampling from effective Hamiltonians
$\overline{\Hc}$ and $\oo{\Hc}$. We denote the corresponding dimer densities by
$\overline{\langle s_1(e_k) \rangle}$ and $\oo{\langle s_1(e_k) \rangle}$
respectively. In Fig.~\ref{fig:compare_effham1}, we compare both these quantities
to $\langle s_1(e_k) \rangle^{\mathrm{ex}}$. We find that while the dimer
densities obtained with $\overline{\Hc}$ accrue a maximum error of  $\sim0.05$,
the ones obtained with $\oo{\Hc}$ have a maximum error of $\sim0.005$. 

\begin{figure}[t]
    \includegraphics[width=\columnwidth]{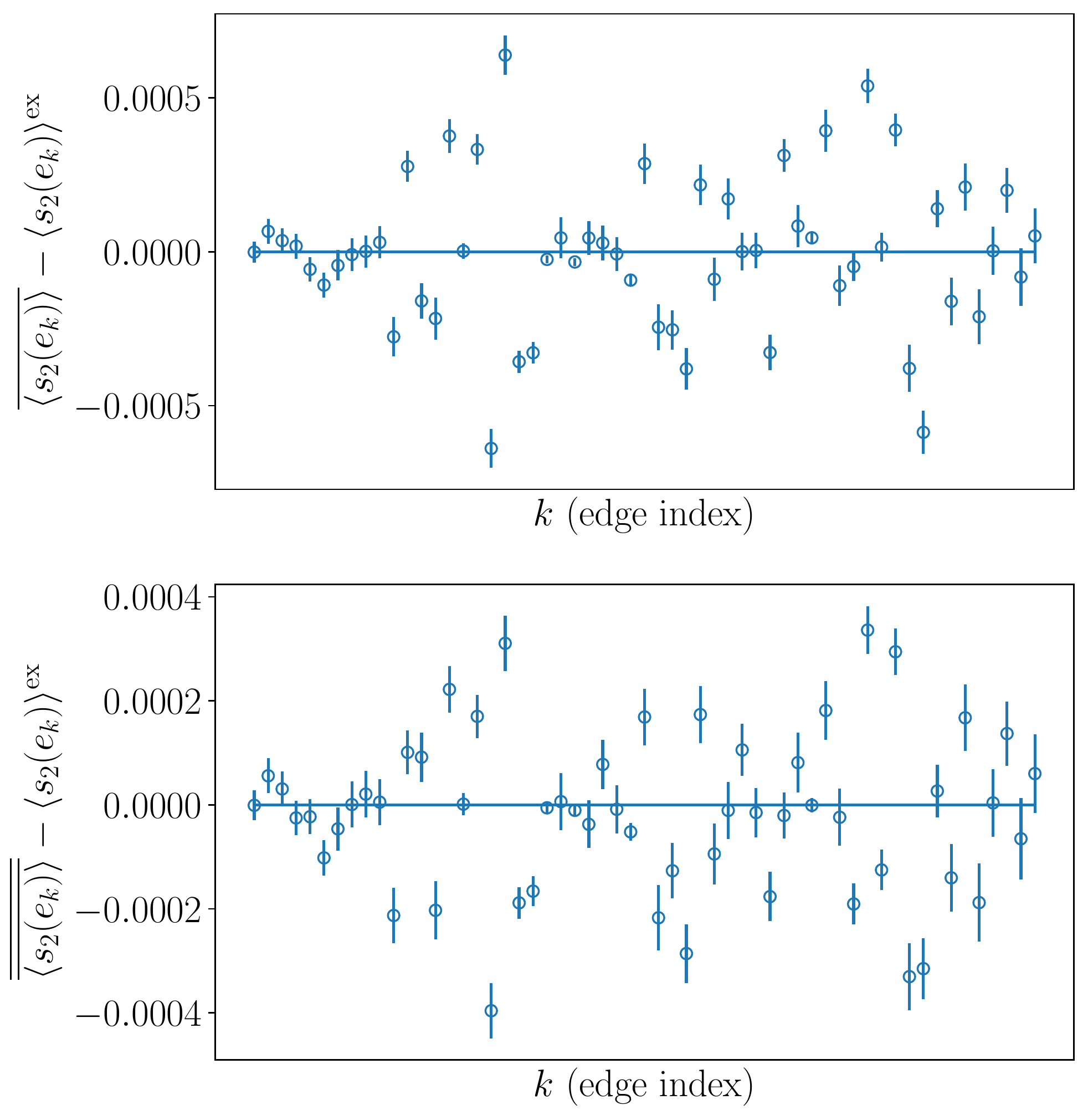}
    \caption{Top (Bottom): The difference of effective dimer-densities $\ovl{\langle s_2(e_k) \rangle}$ ($\oo{\langle s_2(e_k) \rangle}$), calculated with the MCRG coarse-graining of the approximate effective Hamiltonian $\ovl{\Hc}$ ($\oo{\Hc}$)  on an $8_4$-region to an $8_2$-region, to the exact  effective dimer-densities $\langle s_2(e_k)\rangle_{\mathrm{ex}}$, calculated directly from MCRG coarse-graining of the microscopic dimer problem on an $8_6$-region. The errrors, accumulated are very small, suggesting that whatever errors are picked up in our calculations of effective Hamiltonians $\ovl{\Hc}$ and $\oo{\Hc}$ (showing up in errors of dimer-densities in Fig.~\ref{fig:compare_effham1})) are irrelevant, being suppressed under further RG transformations as shown here.}
    \label{fig:compare_effham2}
\end{figure}

While both these approximations might be good enough depending on the purpose,
we argue that the errors incurred by approximating the effective Hamiltonians
with either $\overline{\Hc}$ or $\oo{\Hc}$ are not \emph{relevant} in an RG sense,
\textit{i.e.}, they do not {\it increase} under the RG flow. To see this, we now
exactly compute the effective dimer density  $\langle s_2(e_k)
\rangle^{\mathrm{ex}}$ on the $8_2$-graph, obtained by implementing a ``large"
RG transformation (Eq.~\eqref{eq:rg_transf2})  on dimer configurations on the
$8_6$-graph. We compare this with the densities  $\overline{\langle s_2(e_k)
\rangle}$ and $\oo{\langle s_2(e_k) \rangle}$, obtained by coarse-graining the
effective Hamiltonians $\overline{\Hc}$ and $\oo{\Hc}$ on an $8_4$ graph, the
idea being that this extra  RG transformation might help suppress the irrelevant
parts of any errors incurred in the approximations
$\overline{\Hc}$ and $\oo{\Hc}$ to the effective Hamiltonian in the first step.
The comparisons, displayed in Fig.~\ref{fig:compare_effham2}, are encouraging, in that errors in dimer densities are now upper bounded by $\sim 0.0005$ for both $\overline{\Hc}$ and $\oo{\Hc}$.

Based on this evidence, we proceed with the assumption that the dimer weights
calculated using the algorithm presented at the end of
Sec.~\ref{sec:effham_method}, via the partition function of
Eq.~\eqref{eq:partition_func}, provide a close approximation of effective
Hamiltonians for the purposes of investigating the RG flow of the dimer model
on the \AB~tiling.  To study the RG flow in terms of these effective
Hamiltonians, we start with the microscopic Hamiltonian on an $8_4$-graph where
all dimer weights $W_0(e_k)=1$. Using the algorithm described in
Sec.~\ref{sec:effham_method} we calculate the weights of effective dimers $W_1$
on the coarse-grained $8_2$-graph. Using the approximation $\oo{\Hc}$, which
gives dimer weights of any edge of any \AB~graph in terms of a finite number of
parameters, we now sample this effective Hamiltonian in an $8_4$-graph. This
allows us to calculate the weights $W_2(e_k)$ on an $8_2$-graph--- providing us
with, via the approximation $\oo{\Hc}$, the effective Hamiltonian on the second
RG iteration. This process can be repeated to track the effective Hamiltonian
as a function of the RG iteration. While we have found the choice of approximation between $\oo{\Hc}$ and $\overline{\Hc}$ to not affect much of what follows, 
we parameterize the effective Hamiltonian at each stage in terms of the parameters
$\overline{W_n}(t)$; as before $\overline{W_n}(t)$ is calculated as the mean of calculated weights $W_n(e_k)$ for all edges $e_k$ whose edge-type $T(e_k)$ is equal to $t$.  
This parametrisation is reasonable in light of the fact that the calculated weights $W_n(e_k)$ for edges with same edge-type $t$ are 
narrowly distributed around the mean $\overline{W_n}(t)$ (For $W_1$, this was displayed in Fig.~\ref{fig:edge_weights_type}). The effective Hamiltonian in terms of these parameters as a function of the RG step $n$ is displayed in
Tab.~\ref{tab:effham_flow}. It is evident that our effective Hamiltonians flow
to a fixed point Hamiltonian, the fixed point is stable to whatever errors are picked up by the numerical approximations in this calculation.

\section{Stability of the fixed point to perturbations in parameters of the effective Hamiltonian}
\label{app:dimer_irrelevance}

We use a standard MCRG trick to study at the stability
of the fixed point to simple perturbations in the dimer weights $\ovl{W_n(t)}$~\cite{landau_binder_2021}.  The idea is to consider a
set of couplings $\{K^n_i\}$ describing the effective Hamiltonian at an RG-step
$n$, which presumably flows to a set of fixed-point values $K^*_i$.  For us,
the couplings are given by the logarithm of the dimer weights, $K^n_t \sim \log
\Big(\overline{W_n}(t)\Big)$.  In this Appendix, we approximate the partition
function by the one obtained with an effective Hamiltonian $\overline{\Hc}$ since,
as shown in the previous section, they are a good approximation and they
describe RG flows faithfully.  Under this approximation, the partition function
is given by 
\begin{align} Z &= \exp(\overline{\Hc}^n).  
\end{align}
$\overline{\Hc}^n$ is given in terms of couplings by 
\begin{align} \nonumber
  \overline{\Hc}^n  &= \sum_{e_k} K^n_{T(e_k)} s^{n}(e_k)\\ &= \sum_t K^n_t
  S^n(t).  \label{eq:partition_func2} 
  \end{align} 
  As before, $T(e_k)$ is the
edge-type of  edge $e_k$.  In the last line, the sum is over the ten different edge-types $t$, and we have defined $S^n(t)$ as the
total number of dimers on edges with edge-type $t$. This is the operator to which
$K^n_t$ couples.

To investigate the stability of the fixed point, we calculate the stability
matrix $M^n_{pq}=\cd K^{n+1}_p / \cd K^{n}_q$. This can be done using the `MCRG chain
rule': 
\begin{align} 
  \frac{\cd \langle S^{n+1}_p \rangle }{\cd K^{n}_q}= \sum_r
  \frac{\cd K^{n+1}_r }{\cd K^{n}_q} \frac{\cd \langle S^{n+1}_p \rangle }{\cd
    K^{n+1}_r} \label{eq:mcrg_chainrule}
\end{align}

The advantage of starting from Eq.~\eqref{eq:mcrg_chainrule} is that the derivatives of dimer densities are directly accessible to MC simulations in terms of connected correlation functions :
\begin{align}
  \label{eq:cross_corr}
  \frac{\cd \langle S^{n+1}_p \rangle }{\cd K^{n}_q} &= \la S^{n+1}_p S^{n}_q \ra_c = \la S^{n+1}_p S^{n}_q \ra - \la S^{n+1}_p\ra \la S^{n}_q \ra  \\
  \label{eq:corr}
  \frac{\cd \langle S^{n+1}_p \rangle }{\cd K^{n+1}_q} &= \la S^{n+1}_p S^{n+1}_q \ra_c = \la S^{n+1}_p S^{n+1}_q \ra - \la S^{n+1}_p\ra \la S^{n+1}_q \ra  
\end{align}

Once these correlation functions are computed, the stability matrix 
$M^n_{pq}=\cd K^{n+1}_p / \cd K^{n}_q$ can be computed from Eq.~\eqref{eq:mcrg_chainrule} using
techniques of linear algebra. The eigenvalues of the stability matrix
$\lambda_k$ computed close to a fixed point (large $n$) determine the relevance
of various couplings at the fixed point; $\lambda_k>1$ denote relevant
couplings which grow under RG, while $\lambda_k<1$ denote irrelevant couplings
which go to zero under RG flow. It is conventional to express $\lambda_k=
b^{y_k}$, where $b=\delta_s^2$ is the  length rescaling associated with the
coarse-graining transformation. In this language, $y_k>0$ and $y_k<0$ denote 
relevance and irrelevance respectively.

We consider a dimer model on an $8_4$-region with weights $\overline{W_n}(t)$
for $n=4$ where the system is already very near to the fixed point (See
Tab.~\ref{tab:effham_flow}). We coarse grain this to a dimer model on an
$8_2$-region and compute the stability matrix $M^4_{pq}$ as described above. The eigenvalues computed are $-0.046(1)\pm i 0.048(2), 0.043 \pm i0.032$, $-0.005(4)$, $\pm i0.002$, $0.001$, $0.000$ and $0.000$.  We
find that the eigenvalues are rather small in magnitude. Note
that while imaginary eigenvalues of RG transformations are unusual, they have
been recorded in the literature and often  lead to non-trivial features
like limit cycles in RG flows. In our case, the small real part ensures that
the fixed point is stable to all small perturbations in dimer weights. 
\bibliography{refs}
\end{document}